\documentclass[pra,twocolumn,showpacs]{revtex4-1}
\usepackage{amsmath,amscd,amssymb,color}
\usepackage{graphicx,amsfonts,epsf}
\usepackage{multirow}
\usepackage{epstopdf}
\usepackage{float}
\usepackage{hyperref}
\usepackage{enumerate}
\usepackage{ulem}

\begin{document}
\title{True experimental reconstruction of 
quantum states and processes via convex optimization}
\author{Akshay Gaikwad}
\email{akshayg@iisermohali.ac.in}
\affiliation{Department of Physical Sciences, Indian
Institute of Science Education \& 
Research Mohali, Sector 81 SAS Nagar, 
Manauli PO 140306 Punjab India.}
\author{Arvind}
\email{arvind@iisermohali.ac.in}
\affiliation{Department of Physical Sciences, Indian
Institute of Science Education \& 
Research Mohali, Sector 81 SAS Nagar, 
Manauli PO 140306 Punjab India.}
\author{Kavita Dorai}
\email{kavita@iisermohali.ac.in}
\affiliation{Department of Physical Sciences, Indian
Institute of Science Education \& 
Research Mohali, Sector 81 SAS Nagar, 
Manauli PO 140306 Punjab India.}
\begin{abstract}
We use a constrained convex optimization (CCO) method to
experimentally characterize arbitrary quantum states and
unknown quantum processes on a two-qubit NMR quantum
information processor.  Standard protocols for quantum state
and quantum process tomography are based on linear
inversion, which often result in an unphysical density
matrix and hence an invalid process matrix.  The CCO method
on the other hand, produces  physically valid density
matrices and process matrices, with significantly improved
fidelity as compared to the standard methods.  The
constrained optimization problem is solved with the help of
a semi-definite programming (SDP) protocol.  We use the CCO
method to estimate the Kraus operators and characterize
gates in the presence of errors due to decoherence.  We then
assume Markovian system dynamics and use a Lindblad master
equation in conjunction with the CCO method to completely
characterize the noise processes present in the NMR qubits.
\end{abstract} 
\maketitle 
\section{Introduction}
\label{intro}
Recent decades have seen tremendous advances in research to
engineer high fidelity devices based on quantum
technologies\cite{ladd-nature-2010}.  Characterizing quantum
states and quantum processes in such devices is essential to
evaluating their performance and is typically achieved via
quantum state tomography
(QST)~\cite{james-pra-2001,long-job-2001} and quantum
process tomography (QPT)\cite{obrien-prl-04,chuang-jmo-09}
protocols.  QST and QPT are statistical processes which
comprise two basic elements\cite{bartkiewicz-sr-2016}: (1) a
set of measurements and 2) an estimator which maps the
outcomes of the measurements to an estimate of the unknown
state or process.  Since the ensemble size is finite and
systematic errors are inevitable, there is always some
ambiguity associated with the estimation of an
experimentally created state, which often leads to an
unphysical density
matrix~\cite{miranowicz-pra-2014,Wolk-njp-2019}.  It is
hence imperative to design efficient QST and QPT protocols which
result in physically valid density matrices.

Several tomography protocols have been proposed for both
finite- and infinite-dimensional systems, mainly based on
the least-squares linear inversion
method\cite{xin-prl-2017,li-pra-2017,miranowicz-prb-2015}.
They have been successfully demonstrated on various physical
systems such as nuclear spin ensembles~\cite{vind-pra-2014}
and  photon polarization states~\cite{qi-quantum-inf-2017}.
Several estimation strategies for QST have been proposed as
alternatives to the standard methods, such as maximum
likelihood estimation (MLE)~\cite{shang-pra-2017}, model
averaging approach~\cite{ferrie-njp-2014}, gradient approach
for self-guided QST~\cite{ferrie-prl-2014} and compressed
sensing QST~\citep{yang-pra-2017}.  Similar protocols have
been proposed for QPT, which include ancilla-assisted
QPT~\cite{altepeter-prl-2003}, simplified
QPT~\cite{kosut-njp-2009}, selective QPT using quantum
2-design states~\cite{perito-pra-2018}, self-consistent
QPT~\cite{merkel-pra-2013}, compressed sensing
QPT~\cite{rodionov-prb-2014}, and adaptive measurement-based
QPT~\cite{pogorelov-pra-2017}.  The experimental
implementations of these QST and QPT protocols include
hardware platforms such as
NMR~\cite{maciel-njp-2015,singh-pla-2016,gaikwad-pra-2018},
superconducting qubits~\cite{neeley-nature-2008}, nitrogen
vacancy centers in diamond~\cite{nv-solid,zhang-prl-2014} and
linear optics~\cite{paz-prl-2010,Chapman-prl-2016}.  A
simplified QPT method was developed to experimentally
simulate dephasing channels on an NMR quantum
processor~\cite{wu-jcp-2013}.  All these methods have been
reviewed with respect to their physical resource
requirements and their efficiency~\cite{qpt-review}.

Despite numerous tomography approaches in existence,  most
of them do not produce a valid density or process matrix
after implementation.  On the other hand, protocols such as
adaptive measurements and self-guided tomography which
produce valid states and processes, involve a large number
of projective measurements~\cite{struchalin-pra-2016} which
are experimentally and computationally 
resource-intensive.
In other methods such as the MLE protocol, one needs to {\em
a priori} know the noise distribution present in the
system~\cite{schwemmer-prl-2015}.  In this work, we have
experimentally implemented a method for QST and QPT
that
resolves the issue of the unphysicality of the
experimentally reconstructed density matrix
and process
matrix.  The standard linear inversion based tomography
problem has been transformed into a constrained convex
optimization (CCO) problem~\cite{rabitz-cco,jin-2019}.
The CCO method is based on optimizing a least squares
objective function, subject to the positivity condition as a
nonlinear constraint and the unit trace condition as a
linear constraint. The advantages of the CCO method are
that it does not require any prior knowledge about the
system and does not use extra ancillary qubits. We
demonstrated these advantages of the CCO method by using it
to characterize unknown two-qubit quantum states and
processes on an NMR quantum information processor.
A criterion termed `state deviation' was used to assess how well the
reconstructed quantum process fits the result of the
tomography.  We efficiently computed the complete set of
valid Kraus operators corresponding to a given quantum
process via unitary diagonalization of the experimentally
reconstructed positive process matrix. Finally, a 
Lindbladian approach was used in conjunction with
the  CCO method to study NMR noise
processes inherent in the system.

This paper is organized as follows: In Section~\ref{QST} we
describe the formulation of the CCO problem in the context of
QST, and present experimental results for the
characterization of various two-qubit quantum states.  In
Section~\ref{QPT}, we apply the CCO method to QPT and
describe experiments to characterize several quantum
processes of a two-qubit system.  
In Section~\ref{decohmodel} the CCO
QPT method is used to characterize the noise channels which
are active during decoherence of two NMR qubits.
Section~\ref{compare}
summarizes a comparison of the CCO QPT method with
standard QPT and with simplified QPT methods.
Section~\ref{concl} contains a few concluding remarks.  
The
complete set of Kraus operators corresponding to a given
quantum process, obtained via the CCO method, is given 
in Appendix~\ref{appendix}.
\section{Quantum State Tomography with
Constrained Convex Optimization}  
\label{QST}
Quantum state tomography (QST) is a method to
completely characterize an unknown quantum
state~\cite{james-pra-2001}.  On an ensemble quantum
computer such as NMR, standard QST is carried out by
measuring the expectation values of a fixed set of basis
operators\cite{chuang-rmp-2005}, with  the $n$-qubit density
operator $\rho$ being represented in the tensor product of
the Pauli basis: \begin{equation} \rho =
\sum_{i=0}^{3}\sum_{j=0}^{3}...\sum_{n=0}^{3}
c_{ij...n}\sigma_i \otimes \sigma_j \otimes...\sigma_n
\label{e1} \end{equation} where $c_{00...0} = 1/2^n $,
$\sigma_0$ denotes the $2 \times 2$ identity matrix and
$\sigma_i, i=1,2,3$ are single-qubit Pauli operators.  By
choosing appropriate experimental settings, one can
determine all expectation values
$c_{ij...n}$\cite{singh-pra-2016} and thereby reconstruct
the density matrix.

The standard protocols for QST involve solving linear system
of equations of
the form 
\begin{equation}
Ax=b
\end{equation}
where matrix $A$ is referred to as a fixed coefficient matrix,
the vector $x$ contains elements of the density matrix which needs to 
be reconstructed and
vector $b$ contains actual experimental 
data\cite{long-job-2001}.
One can solve for $x$ by simply inverting the above equation
and a $\rho$ can be reconstructed which is Hermitian and has
unit trace, but there is no guarantee that it will be
positive, since the positivity constraint for a density
matrix to be valid is not explicitly included in the
standard QST protocol.

To always obtain a positive semi-definite density matrix,
the linear inversion-based standard QST problem can hence be
reformulated as a CCO problem using semi-definite
programming (SDP) as follows: \begin{center}
\begin{equation} \begin{aligned} \min_{x} \quad {\Vert Ax-b
\Vert}_2 \\ s.t. \quad \mathrm{Tr}(\rho) = 1 \\ \rho \geq 0
\end{aligned} \label{e6} \end{equation} \end{center} 
The least-squares objective function given in Eq.\ref{e6} is
defined in Reference~\cite{long-job-2001}.  The SDP problem
stated in Eq.\ref{e6} was formulated using the
YALMIP\cite{lofberg-2004} MATLAB package which employs
SeDuMi\cite{sturm-oms-1999} as the SDP solver.  For two
qubits, the objective function has to be optimized over 16
real variables. After solving the SDP problem, a valid
density matrix is obtained from a least squares fit to the
experimental data, which reveals the true quantum state. 

To demonstrate the efficacy of CCO-based QST, we
experimentally prepared and tomographed several two-qubit
quantum states.  All the experiments were performed at room
temperature on an ensemble of $^{13}$C-enriched chloroform
molecules dissolved in acetone-D6 at room temperature on a
Bruker Avance III 600 MHz FT-NMR spectrometer equipped with
a QXI probe. We encoded two qubits using the nuclear spins
$^{1}$H and $^{13}$C.  
The $T_1$ spin-lattice relaxation time for proton and carbon
are found to be 8 sec and 16.5 sec respectively, while the
$T_2$ spin-spin relaxation time for proton and carbon was
measured to be 2.9 sec and 0.3 sec, respectively.
Qubit-selective rf pulses of desired
phase were used to implement local rotation gates; a
$\frac{\pi}{2}$ rf pulse on ${}^{1}$H was of duration 9.4
$\mu $s at a 18.14 W power level, while a $\frac{\pi}{2}$ rf
pulse on ${}^{13}$C was of duration 15.608 $\mu $s at a
179.47 W power level.  The molecular structure, NMR
parameters, state initialization and NMR circuits to achieve
various quantum gates can be found in 
Reference~\cite{gaikwad-pra-2018}.  
\begin{table}[h!]
\caption{
\label{qsteig}
Eigenvalues for the
two-qubit density matrix, obtained from experimentally 
reconstructed  density matrices via standard and CCO QST.}
\begin{tabular}{p{3.7cm}|p{2.4cm}|p{1.45cm}}
\hline
 Quantum state & Standard &  CCO \\
 \hline
 $\vert 00 \rangle$   & 
\begin{tabular}{ll}
-0.0488, &-0.0171,\\
 0.0499, &1.0160
\end{tabular}
& 
\begin{tabular}{l}
0, 0.0225,\\
0, 0.9775
\end{tabular}
\\
\hline
$\vert 01 \rangle$   & 
\begin{tabular}{ll}
-0.0429, &-0.0222,\\
 0.0364, &1.0287
\end{tabular}
& 
\begin{tabular}{l}
0, 0.0067, \\
0, 0.9933
\end{tabular}
\\
\hline
$\vert 10 \rangle$   & 
\begin{tabular}{ll}
-0.1486, &-0.0911, \\
0.1915, &1.0482
\end{tabular}
&
\begin{tabular}{l}
 0, 0.0807, \\0, 0.9193
\end{tabular}
\\
 \hline
 $\vert 11 \rangle$   &
\begin{tabular}{ll}
 -0.1457,& -0.0955,\\ 0.1933,& 1.0480 
\end{tabular}
&
\begin{tabular}{l}
0, 0.0808,\\ 0, 0.9192
\end{tabular}
\\
\hline
$
\frac{1}{\sqrt{2}}
(\vert 00 \rangle + \vert 11 \rangle)$  &
\begin{tabular}{ll}
-0.0822, &-0.0456,\\
 0.0508, &1.0778 
\end{tabular}
  & 
\begin{tabular}{l}
0, 0.0105,\\
0, 0.9895
\end{tabular}
\\
\hline
  $
\frac{1}{\sqrt{2}}
(\vert 01 \rangle - \vert 10 \rangle)$  &
\begin{tabular}{ll}
-0.0950,& -0.0370,\\
 0.0624,& 1.0696 
\end{tabular}
 & 
\begin{tabular}{l}
0, 0.0142,\\
0, 0.9858
\end{tabular}
\\
   \hline
  $
\frac{1}{\sqrt{2}}
(\vert 00 \rangle - \vert 11 \rangle)$  &
\begin{tabular}{ll}
-0.1315, &-0.0455,\\ 
0.1180,& 1.0591
\end{tabular}
& 
\begin{tabular}{l}
0, 0.0592,\\
 0, 0.9408
\end{tabular}
\\
   \hline
  $
\frac{1}{\sqrt{2}}
(\vert 01 \rangle + \vert 10 \rangle)$  &
\begin{tabular}{ll}
-0.1175, &-0.0278,\\
 0.0910, &0.0543
\end{tabular}
  & 
\begin{tabular}{l}
0, 0.0397,\\
 0, 0.9603
\end{tabular}
\\ 
\hline
  $
\frac{1}{\sqrt{2}}
(\vert 01 \rangle + \vert 11 \rangle)$  &
\begin{tabular}{ll}
-0.0892, &-0.0493,\\
 0.1060, &1.0326
\end{tabular}
& 
\begin{tabular}{l}
0, 0.0255,\\ 0, 0.9745
\end{tabular}
\\
  \hline
 $
\frac{1}{\sqrt{2}}
(\vert 00 \rangle + \vert 01 \rangle)$   &
\begin{tabular}{ll}
-0.0587, &-0.0166,\\ 0.0683,& 1.0070
\end{tabular}
& 
\begin{tabular}{l}
0, 0.0375, \\
0, 0.9625
\end{tabular}
 \\
 \hline
  $
\frac{1}{\sqrt{2}}
(\vert 10 \rangle + \vert 11 \rangle)$  &
\begin{tabular}{ll}
-0.1017,& -0.0730,\\
 0.1209,& 1.0538
\end{tabular}
&
\begin{tabular}{l}
 0, 0.0381, \\
0, 0.9619
\end{tabular}
\\
  \hline
   $
\frac{1}{\sqrt{2}}
(\vert 00 \rangle + \vert 10 \rangle)$   &
\begin{tabular}{ll}
-0.0884,& -0.0469,\\
 0.1093,& 1.0260
\end{tabular}
  &
\begin{tabular}{ll}
 0, 0.0303,\\
 0, 0.9697
\end{tabular}
 \\
   \hline
   $
\frac{1}{\sqrt{2}}
(\vert 01 \rangle + i\vert 11 \rangle)$  &
\begin{tabular}{ll}
-0.0936,& -0.0436,\\
 0.0987,& 1.0385 
\end{tabular}
  & 
\begin{tabular}{l}
0, 0.0267, \\0, 0.9733
\end{tabular}
\\
   \hline
   $
\frac{1}{\sqrt{2}}
(\vert 10 \rangle + i\vert 11 \rangle)$   &
\begin{tabular}{ll}
-0.1122,& -0.0962,\\ 
0.1549, &1.0536
\end{tabular}  
  &
\begin{tabular}{l}
 0, 0.0544, \\
0, 0.9456
\end{tabular}
\\
   \hline
   $
\frac{1}{\sqrt{2}}
(\vert 00 \rangle + i\vert 10 \rangle) $   &
\begin{tabular}{ll}
-0.0898,& -0.0420,\\
 0.1028,& 1.0290
\end{tabular}
 & 
\begin{tabular}{l}
0, 0.0304, \\
0, 0.9696
\end{tabular}
\\
\hline
 $
\frac{1}{\sqrt{2}}
(\vert 00 \rangle + i\vert 01 \rangle) $   &
\begin{tabular}{ll}
-0.0862, &-0.0379,\\
 0.0837, &1.0405 
\end{tabular}
 & \begin{tabular}{l}
0, 0.0329,\\ 0, 0.9671 
\end{tabular}
\\
 \hline
 $ \frac{1}{2}(\vert 00 \rangle + \vert 01 \rangle + \vert 10 \rangle +
\vert 11 \rangle)  $   
& 
\begin{tabular}{ll}
-0.0823, &-0.0293,\\
 0.0974, &1.0142
\end{tabular}
& 
\begin{tabular}{l}
0, 0.0293, \\
0, 0.9707
\end{tabular}
\\
 \hline
  $\frac{1}{2} (\vert 00 \rangle + i\vert 01 \rangle + \vert 10 \rangle
+ i\vert 11 \rangle)  $   
& 
\begin{tabular}{ll}
-0.0917, &-0.0619,\\
 0.1120,& 1.0416
\end{tabular}
 & \begin{tabular}{l}0, 0.0298,\\
 0, 0.9702
\end{tabular}
\\
 \hline
 $ \frac{1}{2}(\vert 00 \rangle + \vert 01 \rangle + i\vert 10 \rangle
+i \vert 11 \rangle) $   
&
\begin{tabular}{ll}
 -0.0728, &-0.0110,\\
 0.0770,& 1.0068 \end{tabular}
  & 
\begin{tabular}{l}
0, 0.0298, \\0, 0.9702
\end{tabular}
\\
 \hline
 $ \frac{1}{2}(\vert 00 \rangle + i\vert 01 \rangle + i\vert 10 \rangle
- \vert 11 \rangle) $  
& 
\begin{tabular}{ll}
-0.0828,& -0.0347,\\
 0.0904,& 1.0271
\end{tabular}
& 
\begin{tabular}{l}
0, 0.0234, \\
0, 0.9766
\end{tabular}
\\
\hline
\end{tabular}
\end{table}
The fidelity between the theoretically expected ($\rho^{}_{\rm theo}$) 
and the experimentally reconstructed ($\rho^{}_{\rm expt}$) 
quantum state were computed using the measure\cite{weinstein-prl-2001}:
\begin{equation}
{\mathcal F}(\rho^{}_{\rm expt},\rho^{}_{\rm theo})=
\frac{|{\rm Tr}[\rho^{}_{\rm expt}\rho_{\rm theo}^\dagger]|}
{\sqrt{{\rm Tr}[\rho_{\rm expt}^\dagger\rho^{}_{\rm expt}]
{\rm Tr}[\rho_{\rm theo}^\dagger\rho^{}_{\rm theo}]}}
\label{e7}
\end{equation} 
The fidelities computed using CCO QST for several
different quantum states showed some improvement over those
computed using  standard QST.
However, the main advantage of the CCO QST method is
that the experimentally reconstructed density matrix is
always positive semi-definite 
and
hence always represents a valid quantum state.
The results for various types of states are tabulated in
Table~\ref{qsteig}.
\section{Quantum Process Tomography with 
Constrained Convex Optimization}
\label{QPT}
Quantum process tomography (QPT) 
aims to
characterize an unknown quantum process.  Any quantum state
$\rho$ undergoing a physically valid process can described
by a completely positive (CP) map,  and an unknown process
$\varepsilon$ can be described in the operator-sum
representation~\cite{kraus-book-1983}:
\begin{equation} 
\varepsilon (\rho) = \sum_{i=1}^{d^2} E_i \rho E_i^\dagger 
\label{e8}
\end{equation}
where $E_i$'s are the Kraus operators satisfying $\sum_{i}
E_i E_i^\dagger =
I$. 
The Kraus operators can be expanded 
using a fixed complete set of basis operators $\lbrace A_i
\rbrace $ 
as 
\begin{equation} 
\varepsilon (\rho) = \sum_{m,n=1}^{d^2} \chi_{mn} A_m \rho A_n^\dagger
\label{e9} 
\end{equation} 
where $\chi_{mn} = \sum_{i}a_{im}a_{in}^*$ is called the
process matrix and is a positive Hermitian matrix satisfying the
trace preserving constraint 
$\sum_{m,n} \chi_{mn} A_{n}^{\dagger}
A_m = \cal{I}$~\cite{obrien-prl-04,childs-pra-2001}.
The dimension of the $\chi$ matrix is specified by $n^4-n^2$ parameters for
a Hilbert space of dimension $n$, and hence the computational resources
required for its determination scale exponentially with the number of qubits.
The $\chi$ matrix can be experimentally determined by
preparing a complete set of linearly independent basis operators and
and estimating the output state after the map action and
finally computing all the elements of $\chi_{mn}$ 
from these experimentally estimated output states via
linear equations of the form: 
\begin{equation}
\beta \chi = \lambda
\label{e10}
\end{equation}
where $\beta$ is a coefficient matrix,
vector $\chi$ contains the elements $\lbrace \chi_{mn} \rbrace$ which are to be
determined and vector $\lambda$ is
the experimental data~\cite{childs-pra-2001}. 
Once the $\chi$ matrix is determined, it can be diagonalized
by a unitary transformation $U$ and the Kraus operators can be
determined from this diagonalized $\chi$ matrix using
\begin{equation}
E_i = \sqrt{d_i} \sum_{j} U_{ji} A_j
\end{equation}
where $d_i$ are eigenvalues of $\chi$. 
This
reconstruction of the full set of Kraus operators 
only works if the experimentally determined $\chi$
matrix is positive semidefinite i.e. if the $d_i \ge 0$.

The $\chi$ matrix obtained from standard QPT protocols
is Hermitian and has unit trace, but there is no assurance that it will
be positive. 
Standard QPT methods could hence
lead to an unphysical density matrix which implies that the inversion
was not able to optimally fit the experimental data, and  
more constraints would have to be used
to reconstruct the $\chi$ matrix.  
One viable alternative is the CCO method of 
reconstruction, which always leads to a valid process matrix. 
Convex optimization leads to a global optimization of the
model parameters which best fit the {\it a priori}
information. This circumvents the problem of unphysicality
in standard QPT methods and the genuine action of noise
channels on different input states can be correctly
estimated.
In case of completely positive trace preserving (CPTP)
 maps the mathematical formulation of the CCO
method for QPT is
given by:
\begin{center}
\begin{equation}
\begin{aligned}
\min_{\chi} \quad {\Vert \beta \chi -\lambda \Vert}_2 \\ 
s.t. \quad \sum_{m,n} \chi_{mn} A_{n}^{\dagger}
A_m = \cal{I} \\
     \chi \geq 0
\end{aligned}
\label{e11}
\end{equation}
\end{center} 
The CCO problem given in Eq.~\ref{e11} can be solved efficiently 
using SDP~\cite{lofberg-2004,sturm-oms-1999}. 
For two qubits we used 16 linearly independent density
operators corresponding to quantum states (this choice is
not unique): $\lbrace \vert 00 \rangle$, $\vert 01 \rangle$,
$\vert 0+ \rangle$, $\vert 0- \rangle$, $\vert 10 \rangle$,
$\vert 11 \rangle$, $\vert 1+ \rangle$, $\vert 1- \rangle$,
$\vert +0 \rangle$, $\vert +1 \rangle$, $\vert ++ \rangle$,
$\vert +- \rangle$, $\vert -0 \rangle$, $\vert -1 \rangle$,
$\vert -+ \rangle$ $\vert -- \rangle \rbrace $ where $\vert
+ \rangle = (\vert 0 \rangle + \vert 1 \rangle)/\sqrt{2}$
and $\vert - \rangle = (\vert 0 \rangle + i\vert 1
\rangle)/\sqrt{2}$. 
The dimension of the $\chi$  matrix is $16 \times 16$, the
number of real independent parameters is 255, 
and the vector $\chi$ is of dimensions $256 \times 1$ 
(excluding the trace condition). 
We have to hence optimize 
the objective function over $256$ real variables. 
After solving the SDP problem, we obtain a 
valid $\chi$ matrix, which  can be fitted 
to the experimental data to reveal the true quantum process.
 
\begin{table}[h!] \caption{\label{qpteig} Eigenvalues
obtained from experimental $\chi$ matrices constructed via
standard and CCO QPT.} \begin{tabular}{
p{1.35cm}|p{4.85cm}|p{2.0cm} } \hline Quantum operation&
Standard QPT &  ~CCO QPT\\ \hline CNOT   & 
\begin{tabular}{rrrr}
 1.0117,& 0.1331,& -0.1421,& 0.1247,\\
 0.0934,& 0.0860,& 0.0716,& 0.0541, \\
0.0668,& -0.1135,& -0.0935,& -0.0838,\\
 -0.0315,& -0.0672,& -0.0598,& -0.0503
\end{tabular}
&
\begin{tabular}{l}
0.0077, 0.0201,\\
0.0245, 0.0438,\\0.9038, 0,0,0\\
0,0,0,0,0,0,0,0
\end{tabular}
\\ \hline C-$R^{\pi}_{x}$   &
\begin{tabular}{rrrr}
0.9972,& 0.1435,& -0.1305,& 0.1198,\\
0.1061,& 0.0971,& 0.0837,& 0.0746, \\
0.0553,& -0.0119,& -0.1044,& -0.0838,\\
-0.0415,& -0.0767,& -0.0578,& -0.0639
\end{tabular}
&
\begin{tabular}{l}
0.0077, 0.0166,\\ 0.0315, 0.0397,\\
0.9045, 0,0,0, \\
0,0,0,0,0,0,0,0
\end{tabular}
\\ \hline Identity
& 
\begin{tabular}{rrrr}
1.0087,& 0.1205,& -0.0547,& 0.0581,\\
 0.0355,& -0.0441,& -0.0122,& -0.0385,\\
-0.0338,& -0.0271,& -0.0213,& -0.0151,\\
0.0019,& 0.0006,& -0.0067,& 0.0281
\end{tabular}
&
\begin{tabular}{l}
0.0166, 0.0357, \\
0.9477, 0,0,0,0,\\
0,0,0,0,0,0,0,\\0,0 \\
\end{tabular}
\\ 
\hline \end{tabular} \end{table}

The eigenvalues of experimentally constructed $\chi$
matrices computed via standard and CCO QPT for the
control-$R^{\pi}_{x}$, Identity and CNOT operators are
depicted in Table~\ref{qpteig}. 
As seen from Table~\ref{qpteig}, 
the experimentally estimated $\chi$ matrix via
standard QPT has some negative
eigenvalues which make it unphysical and it does not
correspond to a valid quantum operation. On the other hand,
all the eigenvalues of experimentally estimated $\chi$
matrix via CCO QPT are positive, which makes it physical
and depicts a valid quantum map.

The fidelity of experimentally
constructed $\chi_{\rm{expt}}$ matrix with reference to
theoretically expected $\chi_{\rm{theo}}$ matrix was
calculated using the measure\cite{gaikwad-pra-2018}:
\begin{equation} {\mathcal F}(\chi^{}_{\rm
expt},\chi^{}_{\rm theo})= \frac{|{\rm Tr}[\chi^{}_{\rm
expt}\chi_{\rm theo}^\dagger]|} {\sqrt{{\rm Tr}[\chi_{\rm
expt}^\dagger\chi^{}_{\rm expt}] {\rm Tr}[\chi_{\rm
theo}^\dagger\chi^{}_{\rm theo}]}} \label{e12}
\end{equation} 
The fidelities calculated 
via standard and CCO methods are given in Table~\ref{qptfid}:
In all three cases, the fidelity $\mathcal{F}$ obtained via
CCO method is greater than 0.98, which shows the efficacy of
CCO QPT. 
\begin{table}[h!] \caption{\label{qptfid} Two-qubit gate
fidelities obtained via standard QPT and CCO QPT.}
\begin{tabular}{c c c}
\hline Quantum process &
~~Standard QPT~~& ~~CCO QPT~~\\
\hline Identity & 0.9809 & 0.9959~~~\\ CNOT & 0.9313 &
0.9817~~~\\ control-$R^{\pi}_{x}$  & 0.9269 & 0.9831~~~\\
\hline \end{tabular}
\end{table}

State fidelity cannot be used as a measure of determining
how well the reconstructed process matrix fits the
experimental data, as the first element of the density
matrix dominates the trace.
We 
hence used another metric termed ``Average state
deviation'' $\Delta_{{\rm avg}}$ to 
characterize the quantum process~\cite{jin-2019}:
\begin{equation}
\Delta = \sum_{ij}\frac{(\rm{abs}(\overline{\rho}_{ij}-{\rho_{ij}}))^2}{d^2}
\label{delta}
\end{equation} 
where $\rm{abs}(z)$ denotes the absolute value of complex
number $z$ and $\lbrace \overline{\rho}_{ij} \rbrace$ are
elements of the predicted density matrix using
experimentally constructed $\chi$ matrix while $\lbrace
\rho_{ij} \rbrace $ are elements of ideal gate output.
$\Delta_{{\rm
avg}}$ is then computed by averaging over all the input
states. The smaller the value of $\Delta_{{\rm avg}}$, 
the better the process matrix fits the raw data,
and the
better is the performance of the QPT method.  The average
state deviation $\Delta_{{\rm avg}}$ is given in
Table~\ref{avg}, and it can be seen that the performance of
CCO QPT is much better than standard QPT as $\Delta_{{\rm
avg}}^{{\rm cco}} << \Delta_{{\rm avg}}^{{\rm std}}$  for
all three quantum gates.

\begin{table}[h!]
\caption{\label{avg}
Average state deviation computed from standard 
($\Delta_{{\rm avg}}^{{\rm std}}$) 
and from CCO ($\Delta_{{\rm avg}}^{{\rm cco}}$) methods.}
\begin{tabular}{c c c}
\hline
Quantum process &
~~~~~~$\Delta_{{\rm avg}}^{{\rm std}}$~~~~&
~~~~~~$\Delta_{{\rm avg}}^{{\rm cco}}$~~~~\\
\hline
Identity & 0.0020 & 4.3414e-04~~~\\
CNOT & 0.0097 & 0.0021~~~\\ 
control-$R^{\pi}_{x}$  & 0.0101 & 0.0018~~~\\   
\hline
\end{tabular}
\end{table}
The QPT protocol can be used to
estimate the Kraus operators from the experimental data,
which aid in characterizing the
corresponding quantum gates in presence of various
systematic errors~\cite{nv-solid}.
Three types of errors can occur in the
experimentally constructed density/process matrices: (1)
statistical errors, (2) systematic errors, and (3) errors
due to noisy processes.  To investigate the primary source
of errors for our experimentally constructed density or
process matrices, we numerically simulated the CNOT and
control-$R^{\pi}_{x}$ gates in presence of various
noisy channels~\cite{kofman-pra-2009,childs-pra-2001}.
The complete set of Kraus operators for all three gates are given
in Appendix~\ref{appendix}.
It turns out that the magnitude of extra elements that we
get in the numerically simulated process matrix is of the
order of $10^{-4}$ to $10^{-3}$, while the magnitude of extra
elements of experimentally reconstructed process matrix
using CCO QPT is the order of $10^{-2}$.  This clearly
indicates that the primary source of error in
gate implementation is not
decoherence but rather various systematic errors and
imperfect state preparation due to pulse miscalibration or
rf inhomogeneity~\cite{childs-pra-2001}. 
\begin{figure}
\includegraphics[scale=1]{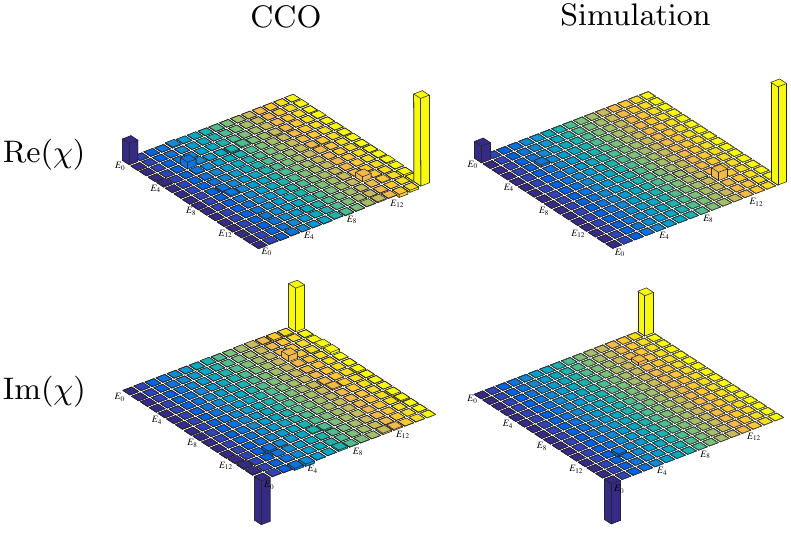}
\caption{Tomographs denoting the real (top
panel) and imaginary (bottom panel) parts of the $\chi$
matrix for system evolving under decoherence for a time
$t=0.05$~s. The tomographs in the first and second columns
represent the experimentally reconstructed $\chi$ matrix
obtained via CCO QPT and via numerical simulation of the
decoherence model.}
\label{fig1}
\end{figure}
\begin{figure}
\includegraphics[scale=1]{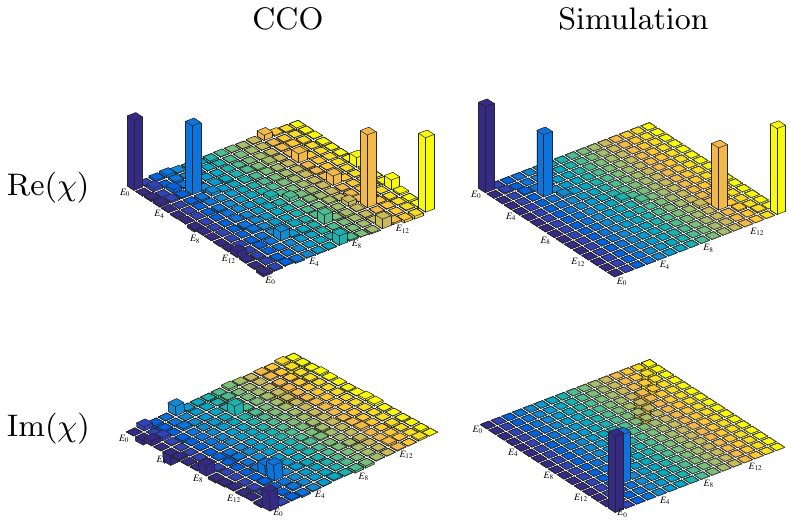}
\caption{Tomographs denoting the real (top
panel) and imaginary (bottom panel) parts of the $\chi$
matrix for system evolving under decoherence for a time
$t=0.5$~s. The tomographs in the first and second columns
represent the experimentally reconstructed $\chi$ matrix
obtained via CCO QPT and via numerical simulation of the
decoherence model.}
\label{fig2}
\end{figure}
\subsection{Markovian Quantum Process Tomography}
\label{decohmodel}
Standard QPT focuses on making predictions about the output
states given an arbitrary set of initial states. However,
the standard method is not able to describe the full system
dynamics. In the regime of Markovian dynamics, one can
construct a valid master equation (called the Lindblad
master equation) which describes time evolution of the
system, via ``snapshots'' of the system captured at
different time points. In such a scenario, the master
equation contains separate terms to describe unitary and
non-unitary evolution~\cite{nv-solid}:
\begin{equation}
\frac{d \rho}{dt} = -i[H,\rho] + \frac{1}{2}
\sum_{k=1}^{d^2-1}
([L_k\rho,L_k^{\dagger}]+[L_k,
\rho L_{k}^{\dagger}])
\end{equation}
where $L_k$ are Lindblad operators describing noise
processes.
\begin{figure}
\includegraphics[scale=1]{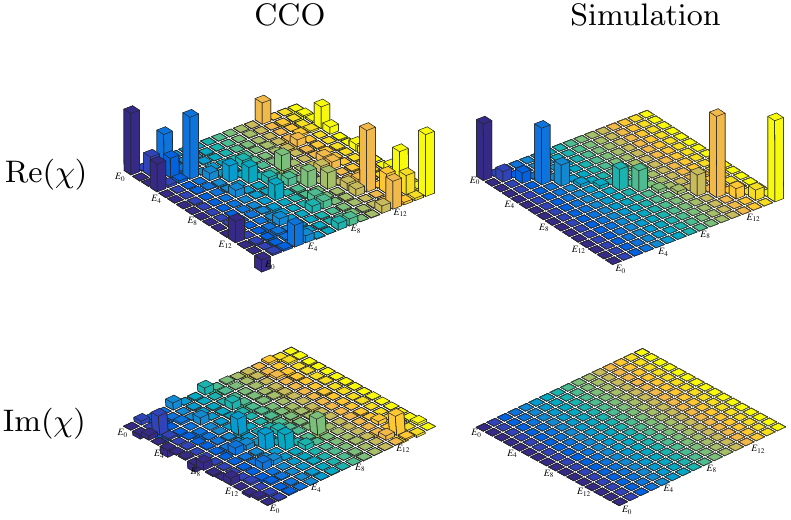}
\caption{Tomographs denoting the real (top
panel) and imaginary (bottom panel) parts of the $\chi$
matrix for system evolving under decoherence for a time
$t=5$~s. The tomographs in the first and second columns
represent the experimentally reconstructed $\chi$ matrix
obtained via CCO QPT and via numerical simulation of the
decoherence model.}
\label{fig3}
\end{figure}

We now proceed to use CCO QPT to characterize the noise
channels acting on the two-qubit NMR system. The relevant
time scales are $1/(2J) \approx 2.33$ msec, $\overline{T}_1
\approx 15$ sec and $\overline{T}_2 \approx 0.5$ sec where
$J$ is scalar spin-spin coupling constant, and
$\overline{T}_1$ and $\overline{T}_2$ are in the range of
the longitudinal ($T_1$) and transverse ($T_2$) relaxation
times. We chose four different time intervals: $t_1 = 0.05$~s,
$t_2 = 0.5$~s, $t_3 = 5$~s and $t_4 = 15$~s, and computed
the $\chi$ matrix for these time points.  
The real and imaginary parts of the tomographed
$\chi$ matrix at the time intervals
$t= 0.05, 0.5, 5$~s are shown in Figs.~\ref{fig1}-~\ref{fig3},
respectively.
We compared our
experimental results to $\chi_{{\rm num}}$ i.e. the $\chi$
matrix obtained by numerically simulating the decoherence
model.  The decoherence model took into account the internal
Hamiltonian of the system, as well as phase damping and
generalized amplitude damping channels acting independently
on each qubit.  We further studied the evolution of
two-qubit maximally entangled Bell states under natural
decoherence using QST and then compared the QST results with
states predicted using CCO QPT as well as those obtained via
numerical simulation of the decoherence model.  To
investigate the goodness of fit of the decoherence model
considered, we calculated process fidelity between
experimentally constructed $\chi$ matrix and the numerically
simulated $\chi_{num}$ for each time point. For the time
intervals $t=0.05$~s, $0.5$~s, $5$~s, and $15$~s, the
calculated fidelities are 0.9901, 0.8441, 0.7245 and 0.6724,
respectively. This implies that, at small time intervals the
process can be modeled well with the decoherence model
considered, whereas at longer time intervals the decoherence
model needs to be modified by including more
terms\cite{kofman-pra-2009}.

We also studied the behavior of the maximally entangled
Bell states: $\vert {\rm B}_1 \rangle = (\vert 00 \rangle +
\vert 11 \rangle)/\sqrt{2}$, $\vert {\rm B}_2 \rangle = (\vert
01 \rangle + \vert 10 \rangle)/\sqrt{2}$, $\vert {\rm B}_3
\rangle = (\vert 00 \rangle - \vert 11 \rangle)/\sqrt{2}$
and $\vert {\rm B}_4 \rangle = (\vert 01 \rangle - \vert 10
\rangle)/\sqrt{2}$, under decoherence. We 
prepared
these states  with 
experimental fidelities of 0.9968,
0.9956, 0.9911 and 0.9942, respectively. The fidelity between
actual evolved state (constructed via CCO QST)
and output state predicted via QPT (both using
the experimental and numerical $\chi$ matrix) is given in
Table~\ref{bell_decoherence}. It is evident
that for short time intervals
(upto $t \approx O(10^{-1})$~s) the decoherence model is
able to predict the dynamics of maximally entangled Bell
states with fidelities $> 0.9$, while CCO QPT is
able to predict the true dynamics on all timescales, with
good fidelity.
\begin{table}[h!]
\caption{\label{bell_decoherence} The fidelity 
difference between an
actual evolved Bell state (denoted by $\vert {\rm B}_i
\rangle$) computed via CCO QPT,
and predicted output state QPT.
The
first column represents the 
different time intervals for which evolution under the
decoherence process was considered.}
\begin{tabular}{ c|c|c|c|c|c }
\hline
Process & & ~~~$\vert {\rm B}_1 \rangle$~~~ 
& ~~~$\vert {\rm B}_2 \rangle$~~~ & 
$~~~\vert {\rm B}_3 \rangle$~~~ & $~~~\vert {\rm B}_4
\rangle~~~$
\\
\hline
\multirow{ 2}{*}{t=0.05 sec} & CCO & 0.9672 & 0.9808 & 0.9767 & 0.9902 \\
& Numerical & 0.9822 & 0.9921 & 0.9844 & 0.9952  \\ \hline
\multirow{ 2}{*}{t=0.5 sec} & CCO & 0.9884 & 0.9899 & 0.9891 & 0.9757 \\
& Numerical & 0.9785 & 0.9795 & 0.9770 & 0.9831 \\ \hline
\multirow{ 2}{*}{t=5 sec} & CCO & 0.9925 & 0.8410 & 0.9946 & 0.8866 \\
& Numerical & 0.6658 & 0.7193 & 0.6642 & 0.7177 \\ \hline
\multirow{ 2}{*}{t=15 sec} & CCO & 0.9964 & 0.9228 & 0.9959 & 0.9031\\
& Numerical & 0.6060 & 0.7121 & 0.6069 & 0.7126 \\ 
\hline
\end{tabular}
\end{table} 
\subsection{Comparison of CCO QPT with Other Protocols}
\label{compare}
More often than not, standard QPT protocols lead to
unphysical density and processes matrices, which is a major
disadvantage. CCO QPT on the other hand, always produces
valid density and process matrices, which represent the true
quantum state and quantum process.  While the experimental
complexity is the same for both the methods, the computed
state fidelities are better via CCO QPT.  The state
deviation obtained via CCO QPT is much smaller than that
obtained via standard QPT, which indicates its better
performance.  CCO based QPT allows us to accurately predict
the operation of a quantum gate on any arbitrary input
state.  The experimentally reconstructed $\chi$ matrix via
CCO QPT allows us to efficiently compute all Kraus
operators, while standard QPT does not even yield valid
Kraus operators.

The simplified QPT protocol~\cite{wu-jcp-2013} requires prior knowledge
about the form of the system-environment interaction which
is in general not possible. However, CCO QPT does not
require any kind of prior knowledge about the
system-environment interaction.  Simplified QPT is not
universal while CCO QPT is universal and is applicable to
any physical system of arbitrary dimensions.  Both methods
produce a valid quantum map and are able to construct all
Kraus operators.
\section{Conclusions}
\label{concl}
In this study, we have used a constrained convex
optimization (CCO) method to completely characterize various
quantum states and quantum processes of two qubits on an NMR
quantum information processor.  Convex optimization is a
search procedure over all operators that satisfies
experimental and mathematical constraints in such a way that
the solutions that emerge are globally optimal.  The results
for QST and QPT tomography using CCO, have been compared
with those obtained using the standard linear
inversion-based methods.  Our experiments demonstrate that
the CCO method produces  physically valid density and
process matrices, which closely resemble the quantum state
being reconstructed or the quantum process whose evolution
is being mapped, respectively.  Furthermore, the fidelities
obtained via the CCO method are higher as compared to the
standard method. We have used the experimentally constructed
process matrix to also compute a complete set of Kraus
operators corresponding to a given quantum process.  

If quantum states are prepared with high fidelity, any
discrepancies between the experimental data and
reconstructed process matrix cannot be attributed to noise.
In such situations, CCO QPT turns out to be a robust method
to investigate the nature of the noise processes present in
the quantum system.  We have assumed system Markovian
dynamics and used the CCO method to characterize the
decoherence processes inherent to the NMR system.  Ongoing
efforts  in our group include using the CCO method to
characterize decoherence present in the system and hence
design targeted state preservation protocols. Our results
are a step forward in the direction of estimating noise and
improving the fidelity of quantum devices.

\begin{acknowledgments}
All the experiments were performed on a Bruker Avance-III
600 MHz FT-NMR spectrometer at the NMR Research Facility of
IISER Mohali. 
\end{acknowledgments}

\begin{thebibliography}{45}%
\makeatletter
\providecommand \@ifxundefined [1]{%
 \@ifx{#1\undefined}
}%
\providecommand \@ifnum [1]{%
 \ifnum #1\expandafter \@firstoftwo
 \else \expandafter \@secondoftwo
 \fi
}%
\providecommand \@ifx [1]{%
 \ifx #1\expandafter \@firstoftwo
 \else \expandafter \@secondoftwo
 \fi
}%
\providecommand \natexlab [1]{#1}%
\providecommand \enquote  [1]{``#1''}%
\providecommand \bibnamefont  [1]{#1}%
\providecommand \bibfnamefont [1]{#1}%
\providecommand \citenamefont [1]{#1}%
\providecommand \href@noop [0]{\@secondoftwo}%
\providecommand \href [0]{\begingroup \@sanitize@url \@href}%
\providecommand \@href[1]{\@@startlink{#1}\@@href}%
\providecommand \@@href[1]{\endgroup#1\@@endlink}%
\providecommand \@sanitize@url [0]{\catcode `\\12\catcode `\$12\catcode
  `\&12\catcode `\#12\catcode `\^12\catcode `\_12\catcode `\%12\relax}%
\providecommand \@@startlink[1]{}%
\providecommand \@@endlink[0]{}%
\providecommand \url  [0]{\begingroup\@sanitize@url \@url }%
\providecommand \@url [1]{\endgroup\@href {#1}{\urlprefix }}%
\providecommand \urlprefix  [0]{URL }%
\providecommand \Eprint [0]{\href }%
\providecommand \doibase [0]{http://dx.doi.org/}%
\providecommand \selectlanguage [0]{\@gobble}%
\providecommand \bibinfo  [0]{\@secondoftwo}%
\providecommand \bibfield  [0]{\@secondoftwo}%
\providecommand \translation [1]{[#1]}%
\providecommand \BibitemOpen [0]{}%
\providecommand \bibitemStop [0]{}%
\providecommand \bibitemNoStop [0]{.\EOS\space}%
\providecommand \EOS [0]{\spacefactor3000\relax}%
\providecommand \BibitemShut  [1]{\csname bibitem#1\endcsname}%
\let\auto@bib@innerbib\@empty
\bibitem [{\citenamefont {Ladd}\ \emph {et~al.}(2010)\citenamefont {Ladd},
  \citenamefont {Jelezko}, \citenamefont {Laflamme}, \citenamefont {Nakamura},
  \citenamefont {Monroe},\ and\ \citenamefont
  {O$^{'}$Brien}}]{ladd-nature-2010}%
  \BibitemOpen
  \bibfield  {author} {\bibinfo {author} {\bibfnamefont {T.~D.}\ \bibnamefont
  {Ladd}}, \bibinfo {author} {\bibfnamefont {F.}~\bibnamefont {Jelezko}},
  \bibinfo {author} {\bibfnamefont {R.}~\bibnamefont {Laflamme}}, \bibinfo
  {author} {\bibfnamefont {Y.}~\bibnamefont {Nakamura}}, \bibinfo {author}
  {\bibfnamefont {C.}~\bibnamefont {Monroe}}, \ and\ \bibinfo {author}
  {\bibfnamefont {J.~L.}\ \bibnamefont {O$^{'}$Brien}},\ }\href
  {https://doi.org/10.1038/nature08812} {\bibfield  {journal} {\bibinfo
  {journal} {Nature}\ }\textbf {\bibinfo {volume} {464}},\ \bibinfo {pages} {45
  EP } (\bibinfo {year} {2010})}\BibitemShut {NoStop}%
\bibitem [{\citenamefont {James}\ \emph {et~al.}(2001)\citenamefont {James},
  \citenamefont {Kwiat}, \citenamefont {Munro},\ and\ \citenamefont
  {White}}]{james-pra-2001}%
  \BibitemOpen
  \bibfield  {author} {\bibinfo {author} {\bibfnamefont {D.~F.~V.}\
  \bibnamefont {James}}, \bibinfo {author} {\bibfnamefont {P.~G.}\ \bibnamefont
  {Kwiat}}, \bibinfo {author} {\bibfnamefont {W.~J.}\ \bibnamefont {Munro}}, \
  and\ \bibinfo {author} {\bibfnamefont {A.~G.}\ \bibnamefont {White}},\ }\href
  {\doibase 10.1103/PhysRevA.64.052312} {\bibfield  {journal} {\bibinfo
  {journal} {Phys. Rev. A}\ }\textbf {\bibinfo {volume} {64}},\ \bibinfo
  {pages} {052312} (\bibinfo {year} {2001})}\BibitemShut {NoStop}%
\bibitem [{\citenamefont {Long}\ \emph {et~al.}(2001)\citenamefont {Long},
  \citenamefont {Yan},\ and\ \citenamefont {Sun}}]{long-job-2001}%
  \BibitemOpen
  \bibfield  {author} {\bibinfo {author} {\bibfnamefont {G.~L.}\ \bibnamefont
  {Long}}, \bibinfo {author} {\bibfnamefont {H.~Y.}\ \bibnamefont {Yan}}, \
  and\ \bibinfo {author} {\bibfnamefont {Y.}~\bibnamefont {Sun}},\ }\href
  {http://stacks.iop.org/1464-4266/3/i=6/a=305} {\bibfield  {journal} {\bibinfo
   {journal} {J Opt B Quantum Semiclassical Opt}\ }\textbf {\bibinfo {volume}
  {3}},\ \bibinfo {pages} {376} (\bibinfo {year} {2001})}\BibitemShut {NoStop}%
\bibitem [{\citenamefont {O$^{'}$Brien}\ \emph {et~al.}(2004)\citenamefont
  {O$^{'}$Brien}, \citenamefont {Pryde}, \citenamefont {Gilchrist},
  \citenamefont {James}, \citenamefont {Langford}, \citenamefont {Ralph},\ and\
  \citenamefont {White}}]{obrien-prl-04}%
  \BibitemOpen
  \bibfield  {author} {\bibinfo {author} {\bibfnamefont {J.~L.}\ \bibnamefont
  {O$^{'}$Brien}}, \bibinfo {author} {\bibfnamefont {G.~J.}\ \bibnamefont
  {Pryde}}, \bibinfo {author} {\bibfnamefont {A.}~\bibnamefont {Gilchrist}},
  \bibinfo {author} {\bibfnamefont {D.~F.~V.}\ \bibnamefont {James}}, \bibinfo
  {author} {\bibfnamefont {N.~K.}\ \bibnamefont {Langford}}, \bibinfo {author}
  {\bibfnamefont {T.~C.}\ \bibnamefont {Ralph}}, \ and\ \bibinfo {author}
  {\bibfnamefont {A.~G.}\ \bibnamefont {White}},\ }\href {\doibase
  10.1103/PhysRevLett.93.080502} {\bibfield  {journal} {\bibinfo  {journal}
  {Phys. Rev. Lett.}\ }\textbf {\bibinfo {volume} {93}},\ \bibinfo {pages}
  {080502} (\bibinfo {year} {2004})}\BibitemShut {NoStop}%
\bibitem [{\citenamefont {Chuang}\ and\ \citenamefont
  {Nielsen}(1997)}]{chuang-jmo-09}%
  \BibitemOpen
  \bibfield  {author} {\bibinfo {author} {\bibfnamefont {I.~L.}\ \bibnamefont
  {Chuang}}\ and\ \bibinfo {author} {\bibfnamefont {M.~A.}\ \bibnamefont
  {Nielsen}},\ }\href {\doibase 10.1080/09500349708231894} {\bibfield
  {journal} {\bibinfo  {journal} {J. Mod. Optics}\ }\textbf {\bibinfo {volume}
  {44}},\ \bibinfo {pages} {2455} (\bibinfo {year} {1997})}\BibitemShut
  {NoStop}%
\bibitem [{\citenamefont {Bartkiewicz}\ \emph {et~al.}(2016)\citenamefont
  {Bartkiewicz}, \citenamefont {{\v C}ernoch}, \citenamefont {Lemr},\ and\
  \citenamefont {Miranowicz}}]{bartkiewicz-sr-2016}%
  \BibitemOpen
  \bibfield  {author} {\bibinfo {author} {\bibfnamefont {K.}~\bibnamefont
  {Bartkiewicz}}, \bibinfo {author} {\bibfnamefont {A.}~\bibnamefont {{\v
  C}ernoch}}, \bibinfo {author} {\bibfnamefont {K.}~\bibnamefont {Lemr}}, \
  and\ \bibinfo {author} {\bibfnamefont {A.}~\bibnamefont {Miranowicz}},\
  }\href {https://doi.org/10.1038/srep19610} {\bibfield  {journal} {\bibinfo
  {journal} {Sci. Rep.}\ }\textbf {\bibinfo {volume} {6}},\ \bibinfo {pages}
  {19610} (\bibinfo {year} {2016})}\BibitemShut {NoStop}%
\bibitem [{\citenamefont {Miranowicz}\ \emph {et~al.}(2014)\citenamefont
  {Miranowicz}, \citenamefont {Bartkiewicz}, \citenamefont
  {Pe\ifmmode~\check{r}\else \v{r}\fi{}ina}, \citenamefont {Koashi},
  \citenamefont {Imoto},\ and\ \citenamefont {Nori}}]{miranowicz-pra-2014}%
  \BibitemOpen
  \bibfield  {author} {\bibinfo {author} {\bibfnamefont {A.}~\bibnamefont
  {Miranowicz}}, \bibinfo {author} {\bibfnamefont {K.}~\bibnamefont
  {Bartkiewicz}}, \bibinfo {author} {\bibfnamefont {J.}~\bibnamefont
  {Pe\ifmmode~\check{r}\else \v{r}\fi{}ina}}, \bibinfo {author} {\bibfnamefont
  {M.}~\bibnamefont {Koashi}}, \bibinfo {author} {\bibfnamefont
  {N.}~\bibnamefont {Imoto}}, \ and\ \bibinfo {author} {\bibfnamefont
  {F.}~\bibnamefont {Nori}},\ }\href {\doibase 10.1103/PhysRevA.90.062123}
  {\bibfield  {journal} {\bibinfo  {journal} {Phys. Rev. A}\ }\textbf {\bibinfo
  {volume} {90}},\ \bibinfo {pages} {062123} (\bibinfo {year}
  {2014})}\BibitemShut {NoStop}%
\bibitem [{\citenamefont {W{\"o}lk}\ \emph {et~al.}(2019)\citenamefont
  {W{\"o}lk}, \citenamefont {Sriarunothai}, \citenamefont {Giri},\ and\
  \citenamefont {Wunderlich}}]{Wolk-njp-2019}%
  \BibitemOpen
  \bibfield  {author} {\bibinfo {author} {\bibfnamefont {S.}~\bibnamefont
  {W{\"o}lk}}, \bibinfo {author} {\bibfnamefont {T.}~\bibnamefont
  {Sriarunothai}}, \bibinfo {author} {\bibfnamefont {G.~S.}\ \bibnamefont
  {Giri}}, \ and\ \bibinfo {author} {\bibfnamefont {C.}~\bibnamefont
  {Wunderlich}},\ }\href {\doibase 10.1088/1367-2630/aaf5f2} {\bibfield
  {journal} {\bibinfo  {journal} {New J. Phys.}\ }\textbf {\bibinfo {volume}
  {21}},\ \bibinfo {pages} {013015} (\bibinfo {year} {2019})}\BibitemShut
  {NoStop}%
\bibitem [{\citenamefont {Xin}\ \emph {et~al.}(2017)\citenamefont {Xin},
  \citenamefont {Lu}, \citenamefont {Klassen}, \citenamefont {Yu},
  \citenamefont {Ji}, \citenamefont {Chen}, \citenamefont {Ma}, \citenamefont
  {Long}, \citenamefont {Zeng},\ and\ \citenamefont {Laflamme}}]{xin-prl-2017}%
  \BibitemOpen
  \bibfield  {author} {\bibinfo {author} {\bibfnamefont {T.}~\bibnamefont
  {Xin}}, \bibinfo {author} {\bibfnamefont {D.}~\bibnamefont {Lu}}, \bibinfo
  {author} {\bibfnamefont {J.}~\bibnamefont {Klassen}}, \bibinfo {author}
  {\bibfnamefont {N.}~\bibnamefont {Yu}}, \bibinfo {author} {\bibfnamefont
  {Z.}~\bibnamefont {Ji}}, \bibinfo {author} {\bibfnamefont {J.}~\bibnamefont
  {Chen}}, \bibinfo {author} {\bibfnamefont {X.}~\bibnamefont {Ma}}, \bibinfo
  {author} {\bibfnamefont {G.}~\bibnamefont {Long}}, \bibinfo {author}
  {\bibfnamefont {B.}~\bibnamefont {Zeng}}, \ and\ \bibinfo {author}
  {\bibfnamefont {R.}~\bibnamefont {Laflamme}},\ }\href {\doibase
  10.1103/PhysRevLett.118.020401} {\bibfield  {journal} {\bibinfo  {journal}
  {Phys. Rev. Lett.}\ }\textbf {\bibinfo {volume} {118}},\ \bibinfo {pages}
  {020401} (\bibinfo {year} {2017})}\BibitemShut {NoStop}%
\bibitem [{\citenamefont {Li}\ \emph {et~al.}(2017)\citenamefont {Li},
  \citenamefont {Huang}, \citenamefont {Luo}, \citenamefont {Li}, \citenamefont
  {Lu},\ and\ \citenamefont {Zeng}}]{li-pra-2017}%
  \BibitemOpen
  \bibfield  {author} {\bibinfo {author} {\bibfnamefont {J.}~\bibnamefont
  {Li}}, \bibinfo {author} {\bibfnamefont {S.}~\bibnamefont {Huang}}, \bibinfo
  {author} {\bibfnamefont {Z.}~\bibnamefont {Luo}}, \bibinfo {author}
  {\bibfnamefont {K.}~\bibnamefont {Li}}, \bibinfo {author} {\bibfnamefont
  {D.}~\bibnamefont {Lu}}, \ and\ \bibinfo {author} {\bibfnamefont
  {B.}~\bibnamefont {Zeng}},\ }\href {\doibase 10.1103/PhysRevA.96.032307}
  {\bibfield  {journal} {\bibinfo  {journal} {Phys. Rev. A}\ }\textbf {\bibinfo
  {volume} {96}},\ \bibinfo {pages} {032307} (\bibinfo {year}
  {2017})}\BibitemShut {NoStop}%
\bibitem [{\citenamefont {Miranowicz}\ \emph {et~al.}(2015)\citenamefont
  {Miranowicz}, \citenamefont {{\"O}zdemir}, \citenamefont {Bajer},
  \citenamefont {Yusa}, \citenamefont {Imoto}, \citenamefont {Hirayama},\ and\
  \citenamefont {Nori}}]{miranowicz-prb-2015}%
  \BibitemOpen
  \bibfield  {author} {\bibinfo {author} {\bibfnamefont {A.}~\bibnamefont
  {Miranowicz}}, \bibinfo {author} {\bibfnamefont {K.}~\bibnamefont
  {{\"O}zdemir}}, \bibinfo {author} {\bibfnamefont {J.}~\bibnamefont {Bajer}},
  \bibinfo {author} {\bibfnamefont {G.}~\bibnamefont {Yusa}}, \bibinfo {author}
  {\bibfnamefont {N.}~\bibnamefont {Imoto}}, \bibinfo {author} {\bibfnamefont
  {Y.}~\bibnamefont {Hirayama}}, \ and\ \bibinfo {author} {\bibfnamefont
  {F.}~\bibnamefont {Nori}},\ }\href {\doibase 10.1103/PhysRevB.92.075312}
  {\bibfield  {journal} {\bibinfo  {journal} {Phys. Rev. B}\ }\textbf {\bibinfo
  {volume} {92}},\ \bibinfo {pages} {075312} (\bibinfo {year}
  {2015})}\BibitemShut {NoStop}%
\bibitem [{\citenamefont {Vind}\ \emph {et~al.}(2014)\citenamefont {Vind},
  \citenamefont {Souza}, \citenamefont {Sarthour},\ and\ \citenamefont
  {Oliveira}}]{vind-pra-2014}%
  \BibitemOpen
  \bibfield  {author} {\bibinfo {author} {\bibfnamefont {F.~A.}\ \bibnamefont
  {Vind}}, \bibinfo {author} {\bibfnamefont {A.~M.}\ \bibnamefont {Souza}},
  \bibinfo {author} {\bibfnamefont {R.~S.}\ \bibnamefont {Sarthour}}, \ and\
  \bibinfo {author} {\bibfnamefont {I.~S.}\ \bibnamefont {Oliveira}},\ }\href
  {\doibase 10.1103/PhysRevA.90.062339} {\bibfield  {journal} {\bibinfo
  {journal} {Phys. Rev. A}\ }\textbf {\bibinfo {volume} {90}},\ \bibinfo
  {pages} {062339} (\bibinfo {year} {2014})}\BibitemShut {NoStop}%
\bibitem [{\citenamefont {Qi}\ \emph {et~al.}(2017)\citenamefont {Qi},
  \citenamefont {Hou}, \citenamefont {Wang}, \citenamefont {Dong},
  \citenamefont {Zhong}, \citenamefont {Li}, \citenamefont {Xiang},
  \citenamefont {Wiseman}, \citenamefont {Li},\ and\ \citenamefont
  {Guo}}]{qi-quantum-inf-2017}%
  \BibitemOpen
  \bibfield  {author} {\bibinfo {author} {\bibfnamefont {B.}~\bibnamefont
  {Qi}}, \bibinfo {author} {\bibfnamefont {Z.}~\bibnamefont {Hou}}, \bibinfo
  {author} {\bibfnamefont {Y.}~\bibnamefont {Wang}}, \bibinfo {author}
  {\bibfnamefont {D.}~\bibnamefont {Dong}}, \bibinfo {author} {\bibfnamefont
  {H.-S.}\ \bibnamefont {Zhong}}, \bibinfo {author} {\bibfnamefont
  {L.}~\bibnamefont {Li}}, \bibinfo {author} {\bibfnamefont {G.-Y.}\
  \bibnamefont {Xiang}}, \bibinfo {author} {\bibfnamefont {H.~M.}\ \bibnamefont
  {Wiseman}}, \bibinfo {author} {\bibfnamefont {C.-F.}\ \bibnamefont {Li}}, \
  and\ \bibinfo {author} {\bibfnamefont {G.-C.}\ \bibnamefont {Guo}},\ }\href
  {\doibase 10.1038/s41534-017-0016-4} {\bibfield  {journal} {\bibinfo
  {journal} {Quant. Inf. Proc.}\ }\textbf {\bibinfo {volume} {3}},\ \bibinfo
  {pages} {19} (\bibinfo {year} {2017})}\BibitemShut {NoStop}%
\bibitem [{\citenamefont {Shang}\ \emph {et~al.}(2017)\citenamefont {Shang},
  \citenamefont {Zhang},\ and\ \citenamefont {Ng}}]{shang-pra-2017}%
  \BibitemOpen
  \bibfield  {author} {\bibinfo {author} {\bibfnamefont {J.}~\bibnamefont
  {Shang}}, \bibinfo {author} {\bibfnamefont {Z.}~\bibnamefont {Zhang}}, \ and\
  \bibinfo {author} {\bibfnamefont {H.~K.}\ \bibnamefont {Ng}},\ }\href
  {\doibase 10.1103/PhysRevA.95.062336} {\bibfield  {journal} {\bibinfo
  {journal} {Phys. Rev. A}\ }\textbf {\bibinfo {volume} {95}},\ \bibinfo
  {pages} {062336} (\bibinfo {year} {2017})}\BibitemShut {NoStop}%
\bibitem [{\citenamefont {Ferrie}(2014{\natexlab{a}})}]{ferrie-njp-2014}%
  \BibitemOpen
  \bibfield  {author} {\bibinfo {author} {\bibfnamefont {C.}~\bibnamefont
  {Ferrie}},\ }\href {http://stacks.iop.org/1367-2630/16/i=9/a=093035}
  {\bibfield  {journal} {\bibinfo  {journal} {New J. Phys.}\ }\textbf {\bibinfo
  {volume} {16}},\ \bibinfo {pages} {093035} (\bibinfo {year}
  {2014}{\natexlab{a}})}\BibitemShut {NoStop}%
\bibitem [{\citenamefont {Ferrie}(2014{\natexlab{b}})}]{ferrie-prl-2014}%
  \BibitemOpen
  \bibfield  {author} {\bibinfo {author} {\bibfnamefont {C.}~\bibnamefont
  {Ferrie}},\ }\href {\doibase 10.1103/PhysRevLett.113.190404} {\bibfield
  {journal} {\bibinfo  {journal} {Phys. Rev. Lett.}\ }\textbf {\bibinfo
  {volume} {113}},\ \bibinfo {pages} {190404} (\bibinfo {year}
  {2014}{\natexlab{b}})}\BibitemShut {NoStop}%
\bibitem [{\citenamefont {Yang}\ \emph {et~al.}(2017)\citenamefont {Yang},
  \citenamefont {Cong}, \citenamefont {Liu}, \citenamefont {Li},\ and\
  \citenamefont {Li}}]{yang-pra-2017}%
  \BibitemOpen
  \bibfield  {author} {\bibinfo {author} {\bibfnamefont {J.}~\bibnamefont
  {Yang}}, \bibinfo {author} {\bibfnamefont {S.}~\bibnamefont {Cong}}, \bibinfo
  {author} {\bibfnamefont {X.}~\bibnamefont {Liu}}, \bibinfo {author}
  {\bibfnamefont {Z.}~\bibnamefont {Li}}, \ and\ \bibinfo {author}
  {\bibfnamefont {K.}~\bibnamefont {Li}},\ }\href {\doibase
  10.1103/PhysRevA.96.052101} {\bibfield  {journal} {\bibinfo  {journal} {Phys.
  Rev. A}\ }\textbf {\bibinfo {volume} {96}},\ \bibinfo {pages} {052101}
  (\bibinfo {year} {2017})}\BibitemShut {NoStop}%
\bibitem [{\citenamefont {Altepeter}\ \emph {et~al.}(2003)\citenamefont
  {Altepeter}, \citenamefont {Branning}, \citenamefont {Jeffrey}, \citenamefont
  {Wei}, \citenamefont {Kwiat}, \citenamefont {Thew}, \citenamefont
  {O$^{'}$Brien}, \citenamefont {Nielsen},\ and\ \citenamefont
  {White}}]{altepeter-prl-2003}%
  \BibitemOpen
  \bibfield  {author} {\bibinfo {author} {\bibfnamefont {J.~B.}\ \bibnamefont
  {Altepeter}}, \bibinfo {author} {\bibfnamefont {D.}~\bibnamefont {Branning}},
  \bibinfo {author} {\bibfnamefont {E.}~\bibnamefont {Jeffrey}}, \bibinfo
  {author} {\bibfnamefont {T.~C.}\ \bibnamefont {Wei}}, \bibinfo {author}
  {\bibfnamefont {P.~G.}\ \bibnamefont {Kwiat}}, \bibinfo {author}
  {\bibfnamefont {R.~T.}\ \bibnamefont {Thew}}, \bibinfo {author}
  {\bibfnamefont {J.~L.}\ \bibnamefont {O$^{'}$Brien}}, \bibinfo {author}
  {\bibfnamefont {M.~A.}\ \bibnamefont {Nielsen}}, \ and\ \bibinfo {author}
  {\bibfnamefont {A.~G.}\ \bibnamefont {White}},\ }\href {\doibase
  10.1103/PhysRevLett.90.193601} {\bibfield  {journal} {\bibinfo  {journal}
  {Phys. Rev. Lett.}\ }\textbf {\bibinfo {volume} {90}},\ \bibinfo {pages}
  {193601} (\bibinfo {year} {2003})}\BibitemShut {NoStop}%
\bibitem [{\citenamefont {Branderhorst}\ \emph {et~al.}(2009)\citenamefont
  {Branderhorst}, \citenamefont {Nunn}, \citenamefont {Walmsley},\ and\
  \citenamefont {Kosut}}]{kosut-njp-2009}%
  \BibitemOpen
  \bibfield  {author} {\bibinfo {author} {\bibfnamefont {M.~P.~A.}\
  \bibnamefont {Branderhorst}}, \bibinfo {author} {\bibfnamefont
  {J.}~\bibnamefont {Nunn}}, \bibinfo {author} {\bibfnamefont {I.~A.}\
  \bibnamefont {Walmsley}}, \ and\ \bibinfo {author} {\bibfnamefont {R.~L.}\
  \bibnamefont {Kosut}},\ }\href
  {http://stacks.iop.org/1367-2630/11/i=11/a=115010} {\bibfield  {journal}
  {\bibinfo  {journal} {New J. Phys.}\ }\textbf {\bibinfo {volume} {11}},\
  \bibinfo {pages} {115010} (\bibinfo {year} {2009})}\BibitemShut {NoStop}%
\bibitem [{\citenamefont {Perito}\ \emph {et~al.}(2018)\citenamefont {Perito},
  \citenamefont {Roncaglia},\ and\ \citenamefont
  {Bendersky}}]{perito-pra-2018}%
  \BibitemOpen
  \bibfield  {author} {\bibinfo {author} {\bibfnamefont {I.}~\bibnamefont
  {Perito}}, \bibinfo {author} {\bibfnamefont {A.~J.}\ \bibnamefont
  {Roncaglia}}, \ and\ \bibinfo {author} {\bibfnamefont {A.}~\bibnamefont
  {Bendersky}},\ }\href {\doibase 10.1103/PhysRevA.98.062303} {\bibfield
  {journal} {\bibinfo  {journal} {Phys. Rev. A}\ }\textbf {\bibinfo {volume}
  {98}},\ \bibinfo {pages} {062303} (\bibinfo {year} {2018})}\BibitemShut
  {NoStop}%
\bibitem [{\citenamefont {Merkel}\ \emph {et~al.}(2013)\citenamefont {Merkel},
  \citenamefont {Gambetta}, \citenamefont {Smolin}, \citenamefont {Poletto},
  \citenamefont {C{\'o}rcoles}, \citenamefont {Johnson}, \citenamefont {Ryan},\
  and\ \citenamefont {Steffen}}]{merkel-pra-2013}%
  \BibitemOpen
  \bibfield  {author} {\bibinfo {author} {\bibfnamefont {S.~T.}\ \bibnamefont
  {Merkel}}, \bibinfo {author} {\bibfnamefont {J.~M.}\ \bibnamefont
  {Gambetta}}, \bibinfo {author} {\bibfnamefont {J.~A.}\ \bibnamefont
  {Smolin}}, \bibinfo {author} {\bibfnamefont {S.}~\bibnamefont {Poletto}},
  \bibinfo {author} {\bibfnamefont {A.~D.}\ \bibnamefont {C{\'o}rcoles}},
  \bibinfo {author} {\bibfnamefont {B.~R.}\ \bibnamefont {Johnson}}, \bibinfo
  {author} {\bibfnamefont {C.~A.}\ \bibnamefont {Ryan}}, \ and\ \bibinfo
  {author} {\bibfnamefont {M.}~\bibnamefont {Steffen}},\ }\href {\doibase
  10.1103/PhysRevA.87.062119} {\bibfield  {journal} {\bibinfo  {journal} {Phys.
  Rev. A}\ }\textbf {\bibinfo {volume} {87}},\ \bibinfo {pages} {062119}
  (\bibinfo {year} {2013})}\BibitemShut {NoStop}%
\bibitem [{\citenamefont {Rodionov}\ \emph {et~al.}(2014)\citenamefont
  {Rodionov}, \citenamefont {Veitia}, \citenamefont {Barends}, \citenamefont
  {Kelly}, \citenamefont {Sank}, \citenamefont {Wenner}, \citenamefont
  {Martinis}, \citenamefont {Kosut},\ and\ \citenamefont
  {Korotkov}}]{rodionov-prb-2014}%
  \BibitemOpen
  \bibfield  {author} {\bibinfo {author} {\bibfnamefont {A.~V.}\ \bibnamefont
  {Rodionov}}, \bibinfo {author} {\bibfnamefont {A.}~\bibnamefont {Veitia}},
  \bibinfo {author} {\bibfnamefont {R.}~\bibnamefont {Barends}}, \bibinfo
  {author} {\bibfnamefont {J.}~\bibnamefont {Kelly}}, \bibinfo {author}
  {\bibfnamefont {D.}~\bibnamefont {Sank}}, \bibinfo {author} {\bibfnamefont
  {J.}~\bibnamefont {Wenner}}, \bibinfo {author} {\bibfnamefont {J.~M.}\
  \bibnamefont {Martinis}}, \bibinfo {author} {\bibfnamefont {R.~L.}\
  \bibnamefont {Kosut}}, \ and\ \bibinfo {author} {\bibfnamefont {A.~N.}\
  \bibnamefont {Korotkov}},\ }\href {\doibase 10.1103/PhysRevB.90.144504}
  {\bibfield  {journal} {\bibinfo  {journal} {Phys. Rev. B}\ }\textbf {\bibinfo
  {volume} {90}},\ \bibinfo {pages} {144504} (\bibinfo {year}
  {2014})}\BibitemShut {NoStop}%
\bibitem [{\citenamefont {Pogorelov}\ \emph {et~al.}(2017)\citenamefont
  {Pogorelov}, \citenamefont {Struchalin}, \citenamefont {Straupe},
  \citenamefont {Radchenko}, \citenamefont {Kravtsov},\ and\ \citenamefont
  {Kulik}}]{pogorelov-pra-2017}%
  \BibitemOpen
  \bibfield  {author} {\bibinfo {author} {\bibfnamefont {I.~A.}\ \bibnamefont
  {Pogorelov}}, \bibinfo {author} {\bibfnamefont {G.~I.}\ \bibnamefont
  {Struchalin}}, \bibinfo {author} {\bibfnamefont {S.~S.}\ \bibnamefont
  {Straupe}}, \bibinfo {author} {\bibfnamefont {I.~V.}\ \bibnamefont
  {Radchenko}}, \bibinfo {author} {\bibfnamefont {K.~S.}\ \bibnamefont
  {Kravtsov}}, \ and\ \bibinfo {author} {\bibfnamefont {S.~P.}\ \bibnamefont
  {Kulik}},\ }\href {\doibase 10.1103/PhysRevA.95.012302} {\bibfield  {journal}
  {\bibinfo  {journal} {Phys. Rev. A}\ }\textbf {\bibinfo {volume} {95}},\
  \bibinfo {pages} {012302} (\bibinfo {year} {2017})}\BibitemShut {NoStop}%
\bibitem [{\citenamefont {Maciel}\ \emph {et~al.}(2015)\citenamefont {Maciel},
  \citenamefont {Vianna}, \citenamefont {Sarthour},\ and\ \citenamefont
  {Oliveira}}]{maciel-njp-2015}%
  \BibitemOpen
  \bibfield  {author} {\bibinfo {author} {\bibfnamefont {T.~O.}\ \bibnamefont
  {Maciel}}, \bibinfo {author} {\bibfnamefont {R.~O.}\ \bibnamefont {Vianna}},
  \bibinfo {author} {\bibfnamefont {R.~S.}\ \bibnamefont {Sarthour}}, \ and\
  \bibinfo {author} {\bibfnamefont {I.~S.}\ \bibnamefont {Oliveira}},\ }\href
  {http://stacks.iop.org/1367-2630/17/i=11/a=113012} {\bibfield  {journal}
  {\bibinfo  {journal} {New J. Phys.}\ }\textbf {\bibinfo {volume} {17}},\
  \bibinfo {pages} {113012} (\bibinfo {year} {2015})}\BibitemShut {NoStop}%
\bibitem [{\citenamefont {Singh}\ \emph
  {et~al.}(2016{\natexlab{a}})\citenamefont {Singh}, \citenamefont {Arvind},\
  and\ \citenamefont {Dorai}}]{singh-pla-2016}%
  \BibitemOpen
  \bibfield  {author} {\bibinfo {author} {\bibfnamefont {H.}~\bibnamefont
  {Singh}}, \bibinfo {author} {\bibnamefont {Arvind}}, \ and\ \bibinfo {author}
  {\bibfnamefont {K.}~\bibnamefont {Dorai}},\ }\href {\doibase
  https://doi.org/10.1016/j.physleta.2016.07.046} {\bibfield  {journal}
  {\bibinfo  {journal} {Phys. Lett. A}\ }\textbf {\bibinfo {volume} {380}},\
  \bibinfo {pages} {3051 } (\bibinfo {year} {2016}{\natexlab{a}})}\BibitemShut
  {NoStop}%
\bibitem [{\citenamefont {Gaikwad}\ \emph {et~al.}(2018)\citenamefont
  {Gaikwad}, \citenamefont {Rehal}, \citenamefont {Singh}, \citenamefont
  {Arvind},\ and\ \citenamefont {Dorai}}]{gaikwad-pra-2018}%
  \BibitemOpen
  \bibfield  {author} {\bibinfo {author} {\bibfnamefont {A.}~\bibnamefont
  {Gaikwad}}, \bibinfo {author} {\bibfnamefont {D.}~\bibnamefont {Rehal}},
  \bibinfo {author} {\bibfnamefont {A.}~\bibnamefont {Singh}}, \bibinfo
  {author} {\bibnamefont {Arvind}}, \ and\ \bibinfo {author} {\bibfnamefont
  {K.}~\bibnamefont {Dorai}},\ }\href {\doibase 10.1103/PhysRevA.97.022311}
  {\bibfield  {journal} {\bibinfo  {journal} {Phys. Rev. A}\ }\textbf {\bibinfo
  {volume} {97}},\ \bibinfo {pages} {022311} (\bibinfo {year}
  {2018})}\BibitemShut {NoStop}%
\bibitem [{\citenamefont {Neeley}\ \emph {et~al.}(2008)\citenamefont {Neeley},
  \citenamefont {Ansmann}, \citenamefont {Bialczak}, \citenamefont {Hofheinz},
  \citenamefont {Katz}, \citenamefont {Lucero}, \citenamefont {O'Connell},
  \citenamefont {Wang}, \citenamefont {Cleland},\ and\ \citenamefont
  {Martinis}}]{neeley-nature-2008}%
  \BibitemOpen
  \bibfield  {author} {\bibinfo {author} {\bibfnamefont {M.}~\bibnamefont
  {Neeley}}, \bibinfo {author} {\bibfnamefont {M.}~\bibnamefont {Ansmann}},
  \bibinfo {author} {\bibfnamefont {R.~C.}\ \bibnamefont {Bialczak}}, \bibinfo
  {author} {\bibfnamefont {M.}~\bibnamefont {Hofheinz}}, \bibinfo {author}
  {\bibfnamefont {N.}~\bibnamefont {Katz}}, \bibinfo {author} {\bibfnamefont
  {E.}~\bibnamefont {Lucero}}, \bibinfo {author} {\bibfnamefont
  {A.}~\bibnamefont {O'Connell}}, \bibinfo {author} {\bibfnamefont
  {H.}~\bibnamefont {Wang}}, \bibinfo {author} {\bibfnamefont {A.~N.}\
  \bibnamefont {Cleland}}, \ and\ \bibinfo {author} {\bibfnamefont {J.~M.}\
  \bibnamefont {Martinis}},\ }\href {\doibase 10.1038/nphys972} {\bibfield
  {journal} {\bibinfo  {journal} {Nature}\ }\textbf {\bibinfo {volume} {4}},\
  \bibinfo {pages} {523} (\bibinfo {year} {2008})}\BibitemShut {NoStop}%
\bibitem [{\citenamefont {Howard}\ \emph {et~al.}(2006)\citenamefont {Howard},
  \citenamefont {Twamley}, \citenamefont {Wittmann}, \citenamefont {Gaebel},
  \citenamefont {Jelezko},\ and\ \citenamefont {Wrachtrup}}]{nv-solid}%
  \BibitemOpen
  \bibfield  {author} {\bibinfo {author} {\bibfnamefont {M.}~\bibnamefont
  {Howard}}, \bibinfo {author} {\bibfnamefont {J.}~\bibnamefont {Twamley}},
  \bibinfo {author} {\bibfnamefont {C.}~\bibnamefont {Wittmann}}, \bibinfo
  {author} {\bibfnamefont {T.}~\bibnamefont {Gaebel}}, \bibinfo {author}
  {\bibfnamefont {F.}~\bibnamefont {Jelezko}}, \ and\ \bibinfo {author}
  {\bibfnamefont {J.}~\bibnamefont {Wrachtrup}},\ }\href {\doibase
  10.1088/1367-2630/8/3/033} {\bibfield  {journal} {\bibinfo  {journal} {New J.
  Phys.}\ }\textbf {\bibinfo {volume} {8}},\ \bibinfo {pages} {33} (\bibinfo
  {year} {2006})}\BibitemShut {NoStop}%
\bibitem [{\citenamefont {Zhang}\ \emph {et~al.}(2014)\citenamefont {Zhang},
  \citenamefont {Souza}, \citenamefont {Brandao},\ and\ \citenamefont
  {Suter}}]{zhang-prl-2014}%
  \BibitemOpen
  \bibfield  {author} {\bibinfo {author} {\bibfnamefont {J.}~\bibnamefont
  {Zhang}}, \bibinfo {author} {\bibfnamefont {A.~M.}\ \bibnamefont {Souza}},
  \bibinfo {author} {\bibfnamefont {F.~D.}\ \bibnamefont {Brandao}}, \ and\
  \bibinfo {author} {\bibfnamefont {D.}~\bibnamefont {Suter}},\ }\href
  {\doibase 10.1103/PhysRevLett.112.050502} {\bibfield  {journal} {\bibinfo
  {journal} {Phys. Rev. Lett.}\ }\textbf {\bibinfo {volume} {112}},\ \bibinfo
  {pages} {050502} (\bibinfo {year} {2014})}\BibitemShut {NoStop}%
\bibitem [{\citenamefont {Schmiegelow}\ \emph {et~al.}(2010)\citenamefont
  {Schmiegelow}, \citenamefont {Larotonda},\ and\ \citenamefont
  {Paz}}]{paz-prl-2010}%
  \BibitemOpen
  \bibfield  {author} {\bibinfo {author} {\bibfnamefont {C.~T.~s.}\
  \bibnamefont {Schmiegelow}}, \bibinfo {author} {\bibfnamefont {M.~A.}\
  \bibnamefont {Larotonda}}, \ and\ \bibinfo {author} {\bibfnamefont {J.~P.}\
  \bibnamefont {Paz}},\ }\href {\doibase 10.1103/PhysRevLett.104.123601}
  {\bibfield  {journal} {\bibinfo  {journal} {Phys. Rev. Lett.}\ }\textbf
  {\bibinfo {volume} {104}},\ \bibinfo {pages} {123601} (\bibinfo {year}
  {2010})}\BibitemShut {NoStop}%
\bibitem [{\citenamefont {Chapman}\ \emph {et~al.}(2016)\citenamefont
  {Chapman}, \citenamefont {Ferrie},\ and\ \citenamefont
  {Peruzzo}}]{Chapman-prl-2016}%
  \BibitemOpen
  \bibfield  {author} {\bibinfo {author} {\bibfnamefont {R.~J.}\ \bibnamefont
  {Chapman}}, \bibinfo {author} {\bibfnamefont {C.}~\bibnamefont {Ferrie}}, \
  and\ \bibinfo {author} {\bibfnamefont {A.}~\bibnamefont {Peruzzo}},\ }\href
  {\doibase 10.1103/PhysRevLett.117.040402} {\bibfield  {journal} {\bibinfo
  {journal} {Phys. Rev. Lett.}\ }\textbf {\bibinfo {volume} {117}},\ \bibinfo
  {pages} {040402} (\bibinfo {year} {2016})}\BibitemShut {NoStop}%
\bibitem [{\citenamefont {Wu}\ \emph {et~al.}(2013)\citenamefont {Wu},
  \citenamefont {Li}, \citenamefont {Zheng}, \citenamefont {Peng},\ and\
  \citenamefont {Feng}}]{wu-jcp-2013}%
  \BibitemOpen
  \bibfield  {author} {\bibinfo {author} {\bibfnamefont {Z.}~\bibnamefont
  {Wu}}, \bibinfo {author} {\bibfnamefont {S.}~\bibnamefont {Li}}, \bibinfo
  {author} {\bibfnamefont {W.}~\bibnamefont {Zheng}}, \bibinfo {author}
  {\bibfnamefont {X.}~\bibnamefont {Peng}}, \ and\ \bibinfo {author}
  {\bibfnamefont {M.}~\bibnamefont {Feng}},\ }\href {\doibase
  10.1063/1.4774119} {\bibfield  {journal} {\bibinfo  {journal} {J. Chem.
  Phys.}\ }\textbf {\bibinfo {volume} {138}},\ \bibinfo {pages} {024318}
  (\bibinfo {year} {2013})}\BibitemShut {NoStop}%
\bibitem [{\citenamefont {Mohseni}\ \emph {et~al.}(2008)\citenamefont
  {Mohseni}, \citenamefont {Rezakhani},\ and\ \citenamefont
  {Lidar}}]{qpt-review}%
  \BibitemOpen
  \bibfield  {author} {\bibinfo {author} {\bibfnamefont {M.}~\bibnamefont
  {Mohseni}}, \bibinfo {author} {\bibfnamefont {A.~T.}\ \bibnamefont
  {Rezakhani}}, \ and\ \bibinfo {author} {\bibfnamefont {D.~A.}\ \bibnamefont
  {Lidar}},\ }\href {\doibase 10.1103/PhysRevA.77.032322} {\bibfield  {journal}
  {\bibinfo  {journal} {Phys. Rev. A}\ }\textbf {\bibinfo {volume} {77}},\
  \bibinfo {pages} {032322} (\bibinfo {year} {2008})}\BibitemShut {NoStop}%
\bibitem [{\citenamefont {Struchalin}\ \emph {et~al.}(2016)\citenamefont
  {Struchalin}, \citenamefont {Pogorelov}, \citenamefont {Straupe},
  \citenamefont {Kravtsov}, \citenamefont {Radchenko},\ and\ \citenamefont
  {Kulik}}]{struchalin-pra-2016}%
  \BibitemOpen
  \bibfield  {author} {\bibinfo {author} {\bibfnamefont {G.~I.}\ \bibnamefont
  {Struchalin}}, \bibinfo {author} {\bibfnamefont {I.~A.}\ \bibnamefont
  {Pogorelov}}, \bibinfo {author} {\bibfnamefont {S.~S.}\ \bibnamefont
  {Straupe}}, \bibinfo {author} {\bibfnamefont {K.~S.}\ \bibnamefont
  {Kravtsov}}, \bibinfo {author} {\bibfnamefont {I.~V.}\ \bibnamefont
  {Radchenko}}, \ and\ \bibinfo {author} {\bibfnamefont {S.~P.}\ \bibnamefont
  {Kulik}},\ }\href {\doibase 10.1103/PhysRevA.93.012103} {\bibfield  {journal}
  {\bibinfo  {journal} {Phys. Rev. A}\ }\textbf {\bibinfo {volume} {93}},\
  \bibinfo {pages} {012103} (\bibinfo {year} {2016})}\BibitemShut {NoStop}%
\bibitem [{\citenamefont {Schwemmer}\ \emph {et~al.}(2015)\citenamefont
  {Schwemmer}, \citenamefont {Knips}, \citenamefont {Richart}, \citenamefont
  {Weinfurter}, \citenamefont {Moroder}, \citenamefont {Kleinmann},\ and\
  \citenamefont {G{\"u}hne}}]{schwemmer-prl-2015}%
  \BibitemOpen
  \bibfield  {author} {\bibinfo {author} {\bibfnamefont {C.}~\bibnamefont
  {Schwemmer}}, \bibinfo {author} {\bibfnamefont {L.}~\bibnamefont {Knips}},
  \bibinfo {author} {\bibfnamefont {D.}~\bibnamefont {Richart}}, \bibinfo
  {author} {\bibfnamefont {H.}~\bibnamefont {Weinfurter}}, \bibinfo {author}
  {\bibfnamefont {T.}~\bibnamefont {Moroder}}, \bibinfo {author} {\bibfnamefont
  {M.}~\bibnamefont {Kleinmann}}, \ and\ \bibinfo {author} {\bibfnamefont
  {O.}~\bibnamefont {G{\"u}hne}},\ }\href {\doibase
  10.1103/PhysRevLett.114.080403} {\bibfield  {journal} {\bibinfo  {journal}
  {Phys. Rev. Lett.}\ }\textbf {\bibinfo {volume} {114}},\ \bibinfo {pages}
  {080403} (\bibinfo {year} {2015})}\BibitemShut {NoStop}%
\bibitem [{\citenamefont {Branderhorst}\ \emph {et~al.}(2008)\citenamefont
  {Branderhorst}, \citenamefont {Walmsley}, \citenamefont {Kosut},\ and\
  \citenamefont {Rabitz}}]{rabitz-cco}%
  \BibitemOpen
  \bibfield  {author} {\bibinfo {author} {\bibfnamefont {M.~P.~A.}\
  \bibnamefont {Branderhorst}}, \bibinfo {author} {\bibfnamefont {I.~A.}\
  \bibnamefont {Walmsley}}, \bibinfo {author} {\bibfnamefont {R.~L.}\
  \bibnamefont {Kosut}}, \ and\ \bibinfo {author} {\bibfnamefont
  {H.}~\bibnamefont {Rabitz}},\ }\href {\doibase 10.1088/0953-4075/41/7/074004}
  {\bibfield  {journal} {\bibinfo  {journal} {J. Phys. B: At. Mol. Opt. Phys.}\
  }\textbf {\bibinfo {volume} {41}},\ \bibinfo {pages} {074004} (\bibinfo
  {year} {2008})}\BibitemShut {NoStop}%
\bibitem [{\citenamefont {Huang}\ \emph {et~al.}(2019)\citenamefont {Huang},
  \citenamefont {Gao}, \citenamefont {Jiao}, \citenamefont {Yan}, \citenamefont
  {Ji},\ and\ \citenamefont {Jin}}]{jin-2019}%
  \BibitemOpen
  \bibfield  {author} {\bibinfo {author} {\bibfnamefont {X.-L.}\ \bibnamefont
  {Huang}}, \bibinfo {author} {\bibfnamefont {J.}~\bibnamefont {Gao}}, \bibinfo
  {author} {\bibfnamefont {Z.-Q.}\ \bibnamefont {Jiao}}, \bibinfo {author}
  {\bibfnamefont {Z.-Q.}\ \bibnamefont {Yan}}, \bibinfo {author} {\bibfnamefont
  {L.}~\bibnamefont {Ji}}, \ and\ \bibinfo {author} {\bibfnamefont {X.-M.}\
  \bibnamefont {Jin}},\ }\href {\doibase
  https://doi.org/10.1016/j.scib.2019.11.009} {\bibfield  {journal} {\bibinfo
  {journal} {Science Bulletin}\ } (\bibinfo {year} {2019}),\
  https://doi.org/10.1016/j.scib.2019.11.009}\BibitemShut {NoStop}%
\bibitem [{\citenamefont {Vandersypen}\ and\ \citenamefont
  {Chuang}(2005)}]{chuang-rmp-2005}%
  \BibitemOpen
  \bibfield  {author} {\bibinfo {author} {\bibfnamefont {L.~M.~K.}\
  \bibnamefont {Vandersypen}}\ and\ \bibinfo {author} {\bibfnamefont {I.~L.}\
  \bibnamefont {Chuang}},\ }\href {\doibase 10.1103/RevModPhys.76.1037}
  {\bibfield  {journal} {\bibinfo  {journal} {Rev. Mod. Phys.}\ }\textbf
  {\bibinfo {volume} {76}},\ \bibinfo {pages} {1037} (\bibinfo {year}
  {2005})}\BibitemShut {NoStop}%
\bibitem [{\citenamefont {Singh}\ \emph
  {et~al.}(2016{\natexlab{b}})\citenamefont {Singh}, \citenamefont {Arvind},\
  and\ \citenamefont {Dorai}}]{singh-pra-2016}%
  \BibitemOpen
  \bibfield  {author} {\bibinfo {author} {\bibfnamefont {A.}~\bibnamefont
  {Singh}}, \bibinfo {author} {\bibnamefont {Arvind}}, \ and\ \bibinfo {author}
  {\bibfnamefont {K.}~\bibnamefont {Dorai}},\ }\href {\doibase
  10.1103/PhysRevA.94.062309} {\bibfield  {journal} {\bibinfo  {journal} {Phys.
  Rev. A}\ }\textbf {\bibinfo {volume} {94}},\ \bibinfo {pages} {062309}
  (\bibinfo {year} {2016}{\natexlab{b}})}\BibitemShut {NoStop}%
\bibitem [{\citenamefont {Lofberg}(2004)}]{lofberg-2004}%
  \BibitemOpen
  \bibfield  {author} {\bibinfo {author} {\bibfnamefont {J.}~\bibnamefont
  {Lofberg}},\ }\href {\doibase 10.1109/CACSD.2004.1393890} {\emph {\bibinfo
  {title} {YALMIP : a toolbox for modeling and optimization in MATLAB}}}\
  (\bibinfo  {publisher} {2004 IEEE International Conference on Robotics and
  Automation (IEEE Cat. No.04CH37508)},\ \bibinfo {year} {2004})\ pp.\ \bibinfo
  {pages} {284--289}\BibitemShut {NoStop}%
\bibitem [{\citenamefont {Sturm}(1999)}]{sturm-oms-1999}%
  \BibitemOpen
  \bibfield  {author} {\bibinfo {author} {\bibfnamefont {J.~F.}\ \bibnamefont
  {Sturm}},\ }\href {\doibase 10.1080/10556789908805766} {\bibfield  {journal}
  {\bibinfo  {journal} {Optimization Methods and Software}\ }\textbf {\bibinfo
  {volume} {11}},\ \bibinfo {pages} {625} (\bibinfo {year} {1999})},\ \Eprint
  {http://arxiv.org/abs/https://doi.org/10.1080/10556789908805766}
  {https://doi.org/10.1080/10556789908805766} \BibitemShut {NoStop}%
\bibitem [{\citenamefont {Weinstein}\ \emph {et~al.}(2001)\citenamefont
  {Weinstein}, \citenamefont {Pravia}, \citenamefont {Fortunato}, \citenamefont
  {Lloyd},\ and\ \citenamefont {Cory}}]{weinstein-prl-2001}%
  \BibitemOpen
  \bibfield  {author} {\bibinfo {author} {\bibfnamefont {Y.~S.}\ \bibnamefont
  {Weinstein}}, \bibinfo {author} {\bibfnamefont {M.~A.}\ \bibnamefont
  {Pravia}}, \bibinfo {author} {\bibfnamefont {E.~M.}\ \bibnamefont
  {Fortunato}}, \bibinfo {author} {\bibfnamefont {S.}~\bibnamefont {Lloyd}}, \
  and\ \bibinfo {author} {\bibfnamefont {D.~G.}\ \bibnamefont {Cory}},\ }\href
  {\doibase 10.1103/PhysRevLett.86.1889} {\bibfield  {journal} {\bibinfo
  {journal} {Phys. Rev. Lett.}\ }\textbf {\bibinfo {volume} {86}},\ \bibinfo
  {pages} {1889} (\bibinfo {year} {2001})}\BibitemShut {NoStop}%
\bibitem [{\citenamefont {Kraus}\ \emph {et~al.}(1983)\citenamefont {Kraus},
  \citenamefont {Bohm}, \citenamefont {Dollard},\ and\ \citenamefont
  {Wootters}}]{kraus-book-1983}%
  \BibitemOpen
  \bibfield  {author} {\bibinfo {author} {\bibfnamefont {K.}~\bibnamefont
  {Kraus}}, \bibinfo {author} {\bibfnamefont {A.}~\bibnamefont {Bohm}},
  \bibinfo {author} {\bibfnamefont {J.}~\bibnamefont {Dollard}}, \ and\
  \bibinfo {author} {\bibfnamefont {W.}~\bibnamefont {Wootters}},\ }\href
  {\doibase 10.1007/3-540-12732-1} {\emph {\bibinfo {title} {States, Effects,
  and Operations: Fundamental Notions of Quantum Theory}}}\ (\bibinfo
  {publisher} {Springer-Verlag Berlin Heidelberg},\ \bibinfo {year}
  {1983})\BibitemShut {NoStop}%
\bibitem [{\citenamefont {Childs}\ \emph {et~al.}(2001)\citenamefont {Childs},
  \citenamefont {Chuang},\ and\ \citenamefont {Leung}}]{childs-pra-2001}%
  \BibitemOpen
  \bibfield  {author} {\bibinfo {author} {\bibfnamefont {A.~M.}\ \bibnamefont
  {Childs}}, \bibinfo {author} {\bibfnamefont {I.~L.}\ \bibnamefont {Chuang}},
  \ and\ \bibinfo {author} {\bibfnamefont {D.~W.}\ \bibnamefont {Leung}},\
  }\href {\doibase 10.1103/PhysRevA.64.012314} {\bibfield  {journal} {\bibinfo
  {journal} {Phys. Rev. A}\ }\textbf {\bibinfo {volume} {64}},\ \bibinfo
  {pages} {012314} (\bibinfo {year} {2001})}\BibitemShut {NoStop}%
\bibitem [{\citenamefont {Kofman}\ and\ \citenamefont
  {Korotkov}(2009)}]{kofman-pra-2009}%
  \BibitemOpen
  \bibfield  {author} {\bibinfo {author} {\bibfnamefont {A.~G.}\ \bibnamefont
  {Kofman}}\ and\ \bibinfo {author} {\bibfnamefont {A.~N.}\ \bibnamefont
  {Korotkov}},\ }\href {\doibase 10.1103/PhysRevA.80.042103} {\bibfield
  {journal} {\bibinfo  {journal} {Phys. Rev. A}\ }\textbf {\bibinfo {volume}
  {80}},\ \bibinfo {pages} {042103} (\bibinfo {year} {2009})}\BibitemShut
  {NoStop}%
\end{thebibliography}

%
\begin{widetext}
\appendix
\section{Kraus operators}
\label{appendix}
The 
complete set of valid Kraus operators 
for the two-qubit system have been
experimentally computed 
using the CCO QPT method. The Kraus operators 
corresponding to the Identity, CNOT gate and control-$R^{\pi}_{x}$ 
gate are given below:
\begin{itemize}
\item Kraus operators corresponding to Identity gate
\[
E_1= {\begin{bmatrix}
    -0.0308 + 0.0457i  & -0.0028 - 0.0077i &  0.0626 + 0.1056i &  0.0022 + 0.0078i \\
   0.0070 + 0.0095i & -0.0393 + 0.0633i &  -0.0055 - 0.0060i &  0.0550 + 0.1068i \\
   0.0755 - 0.0678i &  0.0203 + 0.0042i &  0.0279 - 0.0575i &  0.0052 + 0.0001i \\
   0.0395 + 0.0187i &  0.0850 + 0.0079i &  0.0153 + 0.0005i &  0.0451 - 0.0399i
  \end{bmatrix}}
\]
\[
E_2={ \begin{bmatrix}
    0.0571 + 0.0943i &  -0.0133 + 0.0201i & -0.1932 - 0.0604i & -0.0069 - 0.0085i \\
  -0.0071 + 0.0271i & -0.0082 + 0.0968i &  0.0173 - 0.0019i & -0.1724 - 0.0574i \\
   0.0189 - 0.1154i &  0.0269 - 0.0103i & -0.0281 - 0.1005i &  0.0091 - 0.0038i \\
   0.0481 - 0.0087i &  0.0723 - 0.0485i & -0.0040 + 0.0067i & -0.0220 - 0.0980i
  \end{bmatrix}}
\]
\[
E_3={ \begin{bmatrix}
    -0.0442 - 0.9758i & -0.0014 + 0.0418i &  0.0103 + 0.0259i & -0.0100 - 0.0007i \\
   0.0005 - 0.0271i & -0.0550 - 0.9813i &  0.0095 + 0.0029i &  0.0129 + 0.0272i \\
   0.0101 - 0.0239i &  0.0088 - 0.0002i & -0.0190 - 0.9617i &  0.0233 + 0.0404i \\
  -0.0096 + 0.0017i &  0.0101 - 0.0205i &  0.0242 - 0.0412i &  0.0021 - 0.9671i \\
  \end{bmatrix}}
\]
\item Kraus operators corresponding to CNOT gate
\[
E_1={\begin{bmatrix}
    0.0344 - 0.0042i &  0.0389 + 0.0130i & -0.0068 + 0.0035i & -0.0691 - 0.0003i \\
  -0.0039 - 0.0038i & -0.0208 - 0.0054i & -0.0494 + 0.0543i &  0.0125 + 0.0320i \\
   0.0548 + 0.0194i & -0.0023 - 0.0251i & -0.0714 + 0.0137i & -0.0094 - 0.0117i \\
   0.0208 + 0.0094i &  0.0654 + 0.0264i & -0.0162 + 0.0148i &  0.0221 + 0.0124i 
  \end{bmatrix}}
\]
\[
E_2={\begin{bmatrix}
   0.0124 + 0.0245i &  0.0065 - 0.0008i & -0.0508 - 0.1283i & -0.0079 + 0.0205i \\
   0.0727 + 0.0709i & -0.0132 - 0.0155i & -0.0941 + 0.0168i &  0.0494 - 0.0326i \\
   0.0323 - 0.0020i & -0.0552 + 0.0537i &  0.1017 + 0.0139i & -0.0577 - 0.0248i \\
   0.0400 + 0.0811i & -0.0112 + 0.0281i &  0.0603 + 0.0204i & -0.0381 - 0.0261i
  \end{bmatrix}}
\]
\[
E_3={\begin{bmatrix}
   0.0907 - 0.0140i & -0.0599 + 0.0491i &  0.0581 + 0.0467i &  0.0292 + 0.0058i \\
  -0.0567 + 0.0142i & -0.0978 - 0.0171i &  0.0109 + 0.0093i & -0.0310 - 0.1036i \\
   0.0267 + 0.0135i & -0.0221 + 0.0546i & -0.0700 + 0.0595i & -0.0752 - 0.0404i \\
  -0.0269 - 0.0463i & -0.0340 + 0.0427i &  0.0765 + 0.0205i & -0.1294 + 0.0564i 
  \end{bmatrix}}
\]
\[
E_4={ \begin{bmatrix}
   0.1786 + 0.0344i &  0.1327 - 0.0629i &  0.0228 + 0.0866i & -0.0018 - 0.0201i \\
   0.0052 - 0.0290i & -0.1264 - 0.0353i &  0.0174 - 0.0797i & -0.0397 + 0.0932i \\
  -0.0346 + 0.0199i &  0.0383 - 0.0361i &  0.1008 - 0.0337i & -0.1466 - 0.0222i \\
  -0.0024 - 0.0169i & -0.0415 - 0.0293i &  0.1058 - 0.0050i &  0.1214 - 0.1034i
  \end{bmatrix}}
\]
\[
E_5={\begin{bmatrix}
   0.0706 + 0.9517i & -0.0369 + 0.0847i &  0.0250 + 0.0166i & -0.0245 - 0.0130i \\
  -0.0139 - 0.1052i & -0.1412 + 0.9412i & -0.0442 - 0.0280i & -0.0077 - 0.0040i \\
  -0.0187 + 0.0169i & -0.0218 - 0.0073i & -0.0414 + 0.0215i & -0.0410 + 0.9380i \\
  -0.0065 - 0.0224i & -0.0537 + 0.0269i &  0.0297 + 0.9390i & -0.0516 + 0.0110i
  \end{bmatrix}}
\]

\item Kraus operators corresponding to control-$R^{\pi}_{x}$ gate
\[
E_1={ \begin{bmatrix}
   0.0012 + 0.0210i &  0.0165 + 0.0134i & -0.0744 + 0.0001i & -0.0121 - 0.0364i \\
   0.0359 + 0.0089i &  0.0251 - 0.0080i & -0.0654 - 0.0831i & -0.0097 - 0.0251i \\
  -0.0234 + 0.0153i & -0.0354 - 0.0211i & -0.0461 + 0.0131i &  0.0004 + 0.0068i \\
   0.0365 + 0.0369i & -0.0233 + 0.0167i & -0.0068 + 0.0143i &  0.0125 - 0.0159i
  \end{bmatrix}}
\]
\[
E_2={ \begin{bmatrix}
   0.0153 + 0.0286i &  0.0001 - 0.1142i & -0.0595 - 0.0245i &  0.0153 - 0.0085i \\
  -0.0141 + 0.0290i & -0.0039 - 0.0323i & -0.0097 + 0.0340i &  0.1150 + 0.0166i \\
   0.0058 + 0.0004i & -0.0092 - 0.0963i &  0.0556 + 0.0204i & -0.0167 - 0.0196i \\
  -0.0192 + 0.0136i & -0.0308 + 0.0298i &  0.0407 + 0.0222i & -0.1054 - 0.0405i
  \end{bmatrix}}
\]
\[
 E_3={ \begin{bmatrix}
   0.1537 + 0.0345i &  0.0717 + 0.0257i &  0.0169 - 0.0805i & -0.0006 - 0.0094i \\
  -0.0074 + 0.0235i & -0.1425 - 0.0207i & -0.0232 + 0.0145i &  0.0001 - 0.0276i \\
   0.0178 + 0.0428i & -0.0375 - 0.0168i &  0.0243 - 0.0328i & -0.0154 + 0.1688i \\
  -0.0239 - 0.0132i & -0.0232 + 0.0146i &  0.0017 - 0.1398i &  0.0512 - 0.0880i
  \end{bmatrix}}
\]
\[
 E_4={ \begin{bmatrix}
   0.0686 - 0.0160i & -0.1101 - 0.0036i & -0.0419 - 0.0764i & -0.0257 - 0.0232i \\
  -0.1221 - 0.0570i & -0.0450 + 0.0133i & -0.0657 + 0.0230i &  0.0491 + 0.0496i \\
   0.0651 + 0.0241i &  0.0643 - 0.0430i & -0.1377 + 0.1614i & -0.0061 + 0.0345i \\
  -0.0239 + 0.0021i & -0.0537 - 0.0392i & -0.0594 - 0.0210i &  0.0213 + 0.1907i
  \end{bmatrix}}
\]
\[
 E_5={\begin{bmatrix}
   0.1841 + 0.9399i & -0.0704 + 0.1026i &  0.0143 + 0.0058i &  0.0012 + 0.0004i \\
  -0.0949 - 0.0906i &  0.0979 + 0.9445i &  0.0084 + 0.0134i & -0.0049 + 0.0210i \\
   0.0077 - 0.0108i & -0.0075 + 0.0042i & -0.0249 + 0.0765i &  0.9336 - 0.0790i \\
  -0.0076 - 0.0216i & -0.0092 - 0.0086i &  0.9304 - 0.0835i &  0.0338 + 0.0811i
  \end{bmatrix}}
\]
\end{itemize}
\end{widetext}

\end{document}